\documentclass{aa}  
\usepackage{multirow}
\usepackage{xcolor}
\usepackage{graphicx}
\usepackage{comment}
\usepackage[para,online,flushleft]{threeparttable}
\usepackage{caption}
\usepackage{subcaption}
\usepackage{adjustbox}
\usepackage{gensymb}
\usepackage{overpic} % Allows overlaying text on images
\usepackage{natbib,twoopt}
\usepackage[hyphenbreaks]{breakurl}
\usepackage[breaklinks]{hyperref}      %% to avoid \citeads line fills, add "draft" 
\bibpunct{(}{)}{;}{a}{}{,}             %% natbib format for A&A and ApJ
\definecolor{cobalt}{rgb}{0.06, 0.2, 0.65}
\hypersetup{
  colorlinks,
  citecolor=cobalt,
  linkcolor=[rgb]{0.8, 0.2, 1.0},
  urlcolor=cobalt,
}
\makeatletter
  \newcommandtwoopt{\citeads}[3][][]{\href{http://adsabs.harvard.edu/abs/#3}%
    {\def\hyper@linkstart##1##2{}%
     \let\hyper@linkend\@empty\citealp[#1][#2]{#3}}}
  \newcommandtwoopt{\citepads}[3][][]{\href{http://adsabs.harvard.edu/abs/#3}%
    {\def\hyper@linkstart##1##2{}%
     \let\hyper@linkend\@empty\citep[#1][#2]{#3}}}
  \newcommandtwoopt{\citetads}[3][][]{\href{http://adsabs.harvard.edu/abs/#3}%
    {\def\hyper@linkstart##1##2{}%
     \let\hyper@linkend\@empty\citet[#1][#2]{#3}}}
  \newcommandtwoopt{\citeyearads}[3][][]%
    {\href{http://adsabs.harvard.edu/abs/#3}
    {\def\hyper@linkstart##1##2{}%
     \let\hyper@linkend\@empty\citeyear[#1][#2]{#3}}}
\makeatother

\def\gs{\mathrel{\raise0.35ex\hbox{$\scriptstyle >$}\kern-0.6em \lower0.40ex\hbox{{$\scriptstyle \sim$}}}}
\def\ls{\mathrel{\raise0.35ex\hbox{$\scriptstyle <$}\kern-0.6em \lower0.40ex\hbox{{$\scriptstyle \sim$}}}}
\usepackage{txfonts}
\usepackage[switch]{lineno}
\begin{document} 
  
\title{Revealing the hidden cosmic feast: A z=4.3 galaxy group hosting two optically dark, efficiently star-forming galaxies}

   \titlerunning{A compact galaxy group at $z=4.3$}

   \author{Malte Brinch\inst{1,2,3}
   \and
   Shuowen Jin\inst{1,2,\thanks{Marie Curie Fellow}}
   \and 
   Raphael Gobat\inst{4}
   \and 
   Nikolaj B. Sillassen\inst{1,2}
   \and
   Hiddo Algera\inst{5,6,7}
   \and
   Steven Gillman\inst{1,2}
   \and
   Thomas R. Greve\inst{1,2}
   \and
   Carlos Gomez-Guijarro\inst{8}
   \and
   Bitten Gullberg\inst{1,2}
   \and 
   Jacqueline Hodge\inst{9}
   \and
   Minju Lee\inst{1,2,^{\star}}
   \and
   Daizhong Liu\inst{10}
   \and
   Georgios Magdis\inst{1,2} 
   \and
   Francesco Valentino\inst{1,11}
   }
   \institute{Cosmic Dawn Center (DAWN), Copenhagen, Denmark\\
              \email{malte.brinch@uv.cl}
         \and
             DTU Space, Technical University of Denmark, Elektrovej 327, 2800 Kgs.~Lyngby, Denmark\\ 
             \email{shuji@dtu.dk}
         \and
         Instituto de Física y Astronomía, Universidad de Valparaíso, Avda. Gran Bretana˜ 1111, Valparaíso, Chile
         \and
          Instituto de Física, Pontificia Universidad Católica de Valparaíso, Casilla 4059, Valparaíso, Chile
         \and
         Institute of Astronomy and Astrophysics, Academia Sinica, 11F of Astronomy-Mathematics Building, No.1, Sec. 4, Roosevelt Rd, Taipei 106216, Taiwan, R.O.C
         \and 
         Hiroshima Astrophysical Science Center, Hiroshima University, 1-3-1 Kagamiyama, Higashi-Hiroshima, Hiroshima 739-8526, Japan
         \and
         National Astronomical Observatory of Japan, 2-21-1, Osawa, Mitaka, Tokyo, Japan
         \and
         AIM, CEA, CNRS, Université Paris-Saclay, Université Paris Diderot, Sorbonne Paris Cité 91191 Gif-sur-Yvette, France
         \and
         Leiden Observatory, Leiden University, NL-2300 RA Leiden, Netherlands
         \and
         Purple Mountain Observatory, Chinese Academy of Sciences, 10 Yuanhua Road, Nanjing 210023, China
         \and
         European Southern Observatory, Karl-Schwarzschild-Str. 2, 85748 Garching, Germany
             }

   \date{Received month day, year; accepted month day, year}
      
% \abstract{}{}{}{}{} 
% 5 {} token are mandatory
 
  \abstract{
  We present the confirmation of a compact galaxy group candidate, CGG-z4, at $z=4.3$ in the COSMOS field. This structure was identified by two spectroscopically confirmed $z=4.3$ $K_s$-dropout galaxies with ALMA $870\rm\, \mu m$ and 3 mm continuum detections, surrounded by an overdensity of near infrared-detected galaxies with consistent photometric redshifts of $4.0<z<4.6$. 
  The two ALMA sources, CGG-z4.a and CGG-z4.b, have been detected with both CO(4-3) and CO(5-4) lines, whereby [CI](1-0) has been detected on CGG-z4.a, and H$_{2}$O($1_{1,0}-1_{0,1}$) absorption detected on CGG-z4.b. We modeled an integrated spectral energy distribution (SED) by combining the far-infrared-to-radio photometry of this group and estimated a total star formation rate of $\rm\sim2000\, M_{\odot}$ yr$^{-1}$,  
  making it one of the most star-forming groups known at $z>4$.
  Their high CO(5-4)/CO(4-3) ratios 
  indicate that each respective interstellar medium (ISM) is close to thermalization, suggesting either high gas temperatures, high densities, and/or high pressure; whereas the low [CI](1-0)/CO(4-3) line ratios indicate high star formation efficiencies.
  With the [CI]-derived gas masses,  
  we found the two galaxies have extremely short gas depletion times of 99 Myr and $<63$ Myr, respectively, suggesting the onset of quenching. 
  With an estimated halo mass of $\rm log (M_{\rm halo}[M_{\odot}])\sim12.8$, we find that this structure is likely to be in the process of forming a massive galaxy cluster.
  }
  \keywords{galaxies: formation – galaxies: evolution – galaxies: high-redshift – galaxies: ISM – galaxies: groups: individual: CGG-z4}
   \maketitle
       
\section{Introduction}
Within the modern $\Lambda$CDM paradigm, a major goal is to understand the formation and evolution of massive galaxies ($\rm M_{\star}>10^{11}\,\rm M_{\odot}$) and the structures they inhabit (clusters and groups). For a $\Lambda$CDM cosmology, massive galaxies are thought to form through the hierarchical clustering of lower mass galaxies and their dark matter halos \citep{Springel2005}. From cosmological simulations, the growth of massive galaxies has been shown to be in a heightened phase prior to cosmic noon ($z>2$) due to a combination of in situ gas accretion and the merging of low-mass ($\rm M_{\star}\lesssim 10^{10}\,\rm M_{\odot}$) satellite galaxies \citep{Hopkins2009,Oser2010,Benson2012,Hirschmann2012}.
Overdense structures such as protoclusters or galaxy groups are prime locations to investigate the growth of massive galaxies, as these structures initially formed in the peaks of the primordial density field. They are thought to undergo a downsizing effect, where the galaxies in the overdense structure (especially in the center of the structure) will become more evolved, compared to the field population \citep{Overzier2016,Chiang2017,Marrone2018}. These overdense structures are typically found through either the use of large-scale surveys of specific types of galaxies such as Lyman-break galaxies or Lyman-alpha emitters \citep[see][]{Harikane2019,Brinch2023,Brinch2024}. They can also be found using signpost galaxies such as submillimeter (submm) galaxies (SMGs) or quasi-stellar objects (QSOs) that are biased tracers of overdensities, where their nearby environment can then be investigated \citep[see][]{Capak2011,Walter2012,Lewis2018,Calvi2023}. 
Wide-field submillimeter-millimeter (submm/mm) and interferometric observations with instruments such as Submillimetre Common-User Bolometer Array 2 (SCUBA-2) on the \textit{James Clark Maxwell} Telescope (JCMT), 
Northern Extended Millimeter Array (NOEMA), and the Atacama Large Millimeter/submillimeter Array (ALMA) have allowed for the study of massive, highly star-forming galaxies in the high-redshift Universe thanks to their high spectral coverage and sensitivity \citep{Dannerbauer2014,Oteo2018,Miller2018}. In conjunction with optical and near-infrared (NIR) surveys, these observations enable us to identify and study the dense and compact structures where these galaxies assemble (e.g., \citealt{Wang2016,Daddi2021,Daddi2022,Jones2017,Diaz-Santos2018,Cooper2022,Ginolfi2022,Sillassen2022,Sillassen2024,Jin2023,Jin2024}).

Meanwhile, studies have revealed a population of optically/NIR dark dusty star-forming galaxies (DSFGs) that drop out in deep optical images \citep{Smail1999, Higdon2008, Wang2019, Algera2020, Smail2021, Talia2021, Guijarro2022, Shu2022, Vlugt2023, Xiao2023}.
They are found to play a critical role in the dust-obscured cosmic star-formation density (SFRD), with a contribution of up to 50\% at high redshift and likely dominate at the massive end ($\rm log(M_{\odot}/M_\odot)>10.5$) of the stellar mass function (SMF) at $z\sim3-7$ \citep{Wang2019}.
With the advent of the James Webb Space Telescope (JWST), The search for optically faint and dark galaxies has been pushed to redshifts of $z\sim8$, where they have been shown to be an important contribution to the high-mass end of the galaxy stellar mass function and the cosmic SFRD up to $z\sim7-8.$  Furthermore, a significant fraction of $z>3$ galaxies, especially massive galaxies, has gone unaccounted for \citep{Barrufet2023,Gottumukkala2024}.
However, the majority of literature studies of this population at high redshift rely heavily on photometric redshifts. The spectroscopically confirmed sample of optically dark galaxies is still small, with their environments and ISM properties remaining largely unknown. 

In this work, we present the discovery of a compact galaxy group (which we dub CGG-z4) and study the member galaxies' star formation and ISM properties. 
This paper is organized as follows.
We describe the observational data and selection in Sect. 2. 
Section 3 describes the methods for redshift confirmation, spectral stacking, and SED fittings. 
In Sect. 4, we present the results and analysis, including different physical properties such as SFR and gas mass estimates. 
We discuss relevant science in Sect. 5 and summarize our conclusions in Sect. 6.
Throughout the paper, we have adopted a standard $\Lambda$CDM cosmology
with $H_{\rm 0}=70\,{\rm km\,s^{-1}\,Mpc^{-1}}$, $\Omega_{\rm m} = 0.3$, 
and $\Omega_{\rm \Lambda} = 0.7$. All magnitudes are expressed in the
AB system \citep{Oke1974}. We adopted a \citet{Chabrier2003} stellar initial mass
function (IMF). The results are reported with uncertainties within the 68\% confidence interval.

%--------------------------------------------------------------------
\section{Selection and data}
\begin{figure*}
    \centering
    \includegraphics[width=0.5113\textwidth]{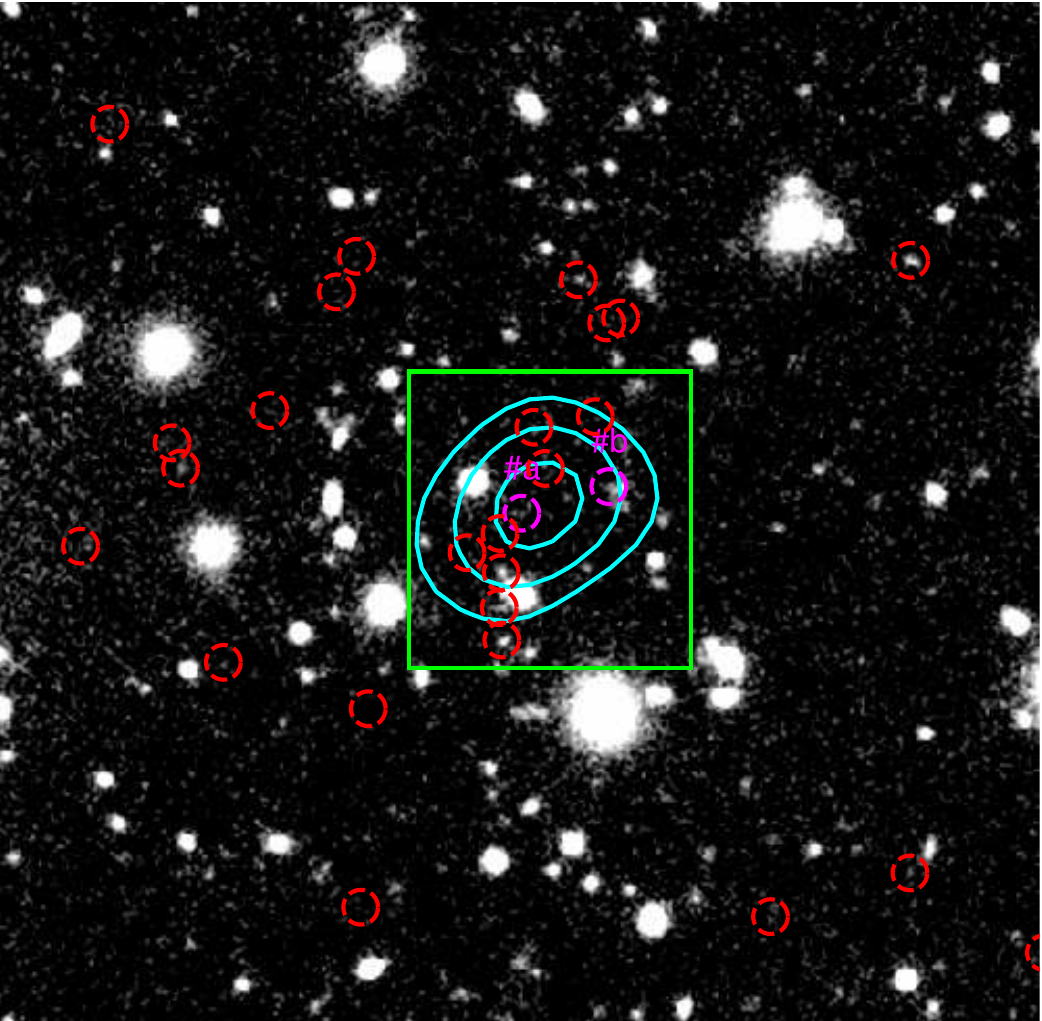}
    \includegraphics[width=0.4758\textwidth]{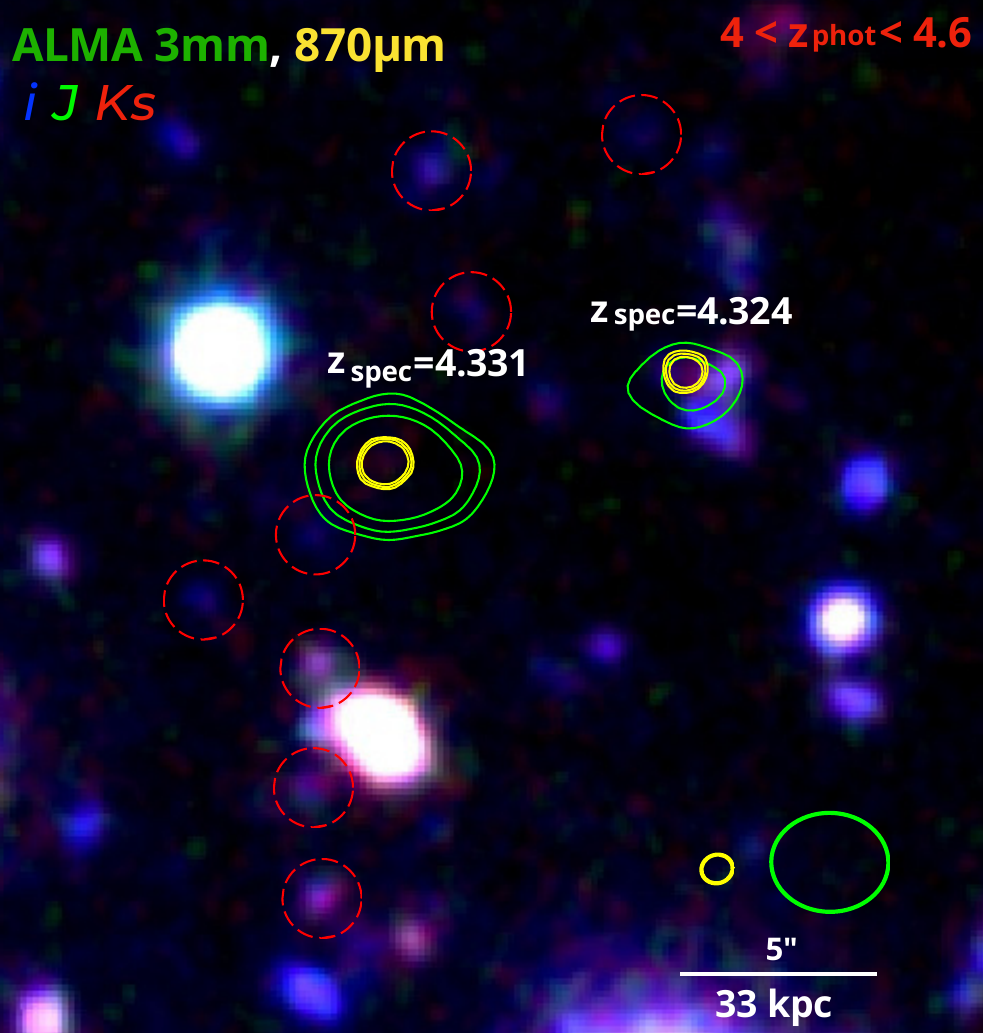}
    \includegraphics[width=\textwidth]{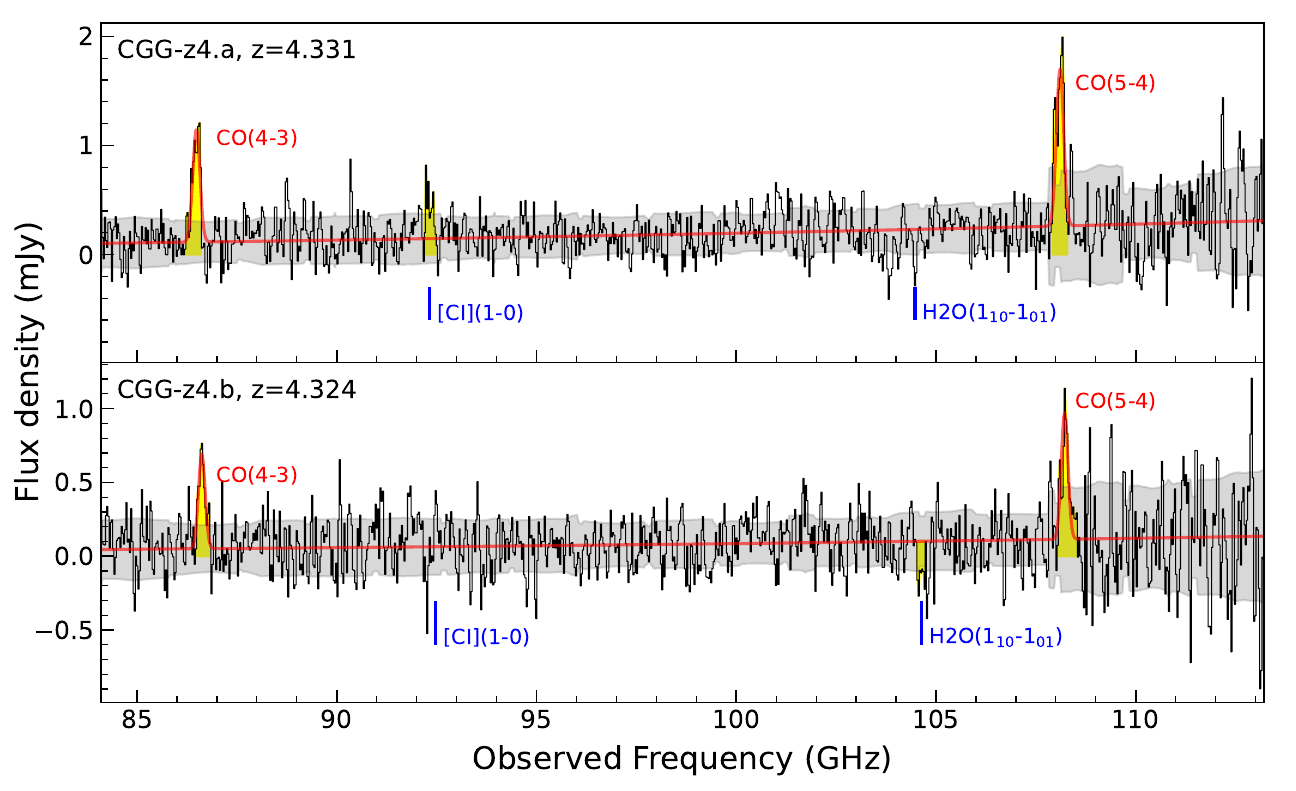}
    \caption{Images and spectra highlighting the member galaxies of the CGG-$z4$ galaxy group. \textbf{Top left:} VISTA Ks images overlaid with cyan SCUBA-2 850$\,\rm \mu m$ contours in steps of 3, 5 and 7$\sigma$. Photometric redshift candidate members in the range $4.0<z<4.6$ are highlighted with red dashed circles. Galaxies with ALMA continuum detections are highlighted as magenta-dashed circles. The green square highlights the region of the right image. \textbf{Top right:} VISTA RGB color images of CGG-z4 at $z\sim4.3$ using the Ks (red), J (green), and i (blue) bands. Two spectroscopically confirmed galaxies have ALMA 3mm and 870$\,\rm \mu m$ continuum emission shown as green and yellow contours. Contours are shown at levels 5, 7, and 10$\sigma$. ALMA beam sizes and an image scale are shown in the lower right corner. The RGB frames are composed using linear scales with identical limits. \textbf{Bottom:} ALMA 3mm spectra for galaxies with emission line detections. The detected lines are highlighted in yellow with their names labeled. The red line shows Gaussian fits to the CO emission lines and a power law fit to the continuum that increases with frequency following the expression $\nu^{3.7}$.}
    \label{fig:color-image}
\end{figure*}

\begin{figure*}
    \centering
    \includegraphics[width=0.36\textwidth]{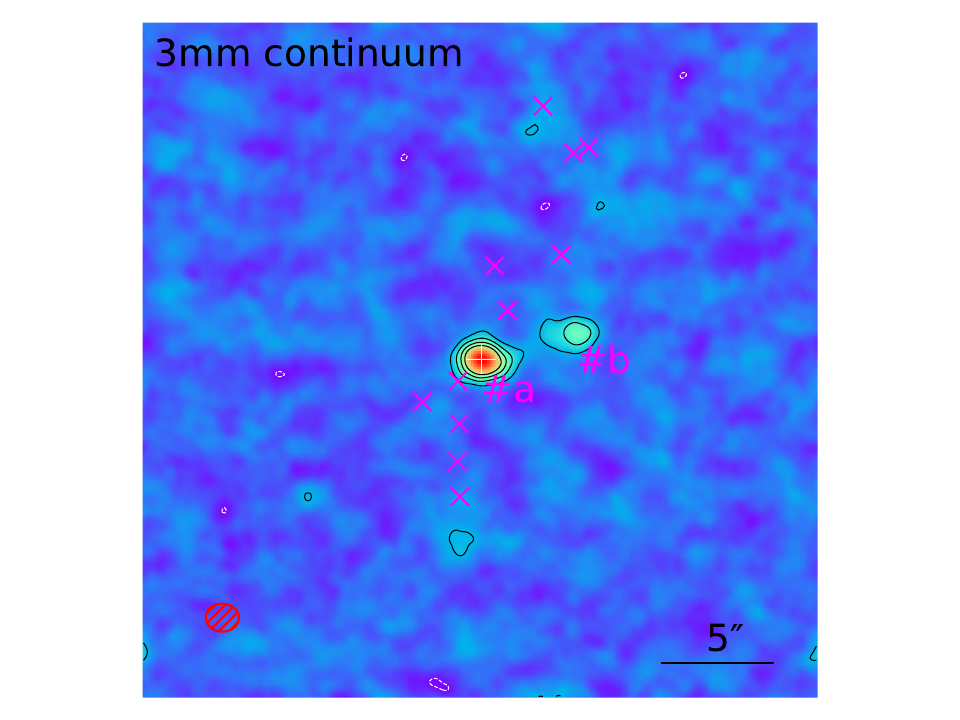}
    \hspace{-1cm}    
    \includegraphics[width=0.36\textwidth]{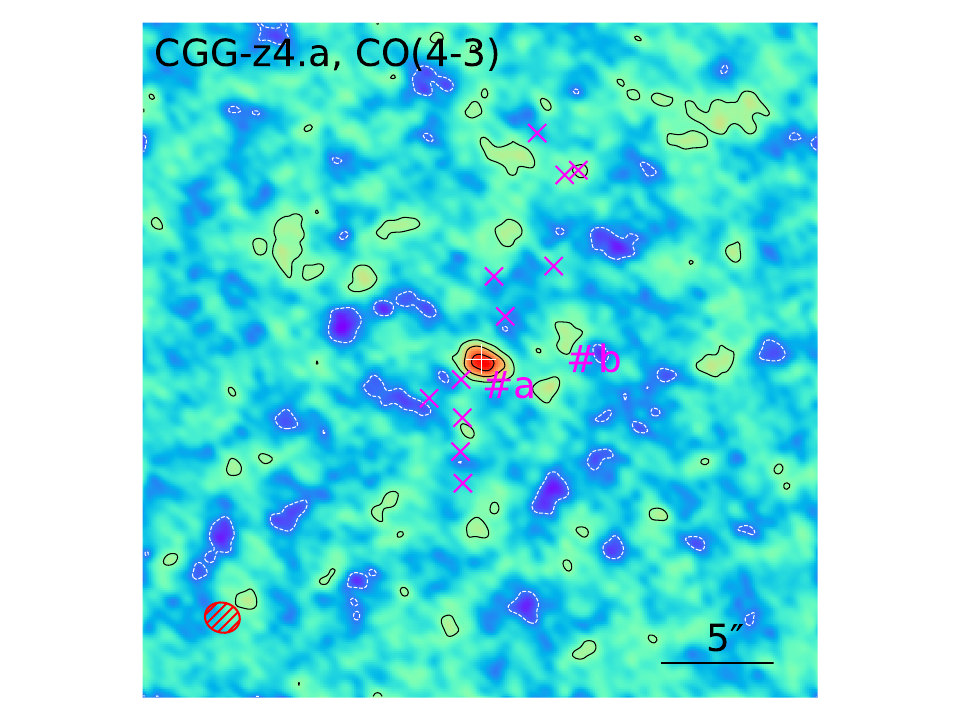}
    \hspace{-1cm}
    \includegraphics[width=0.36\textwidth]{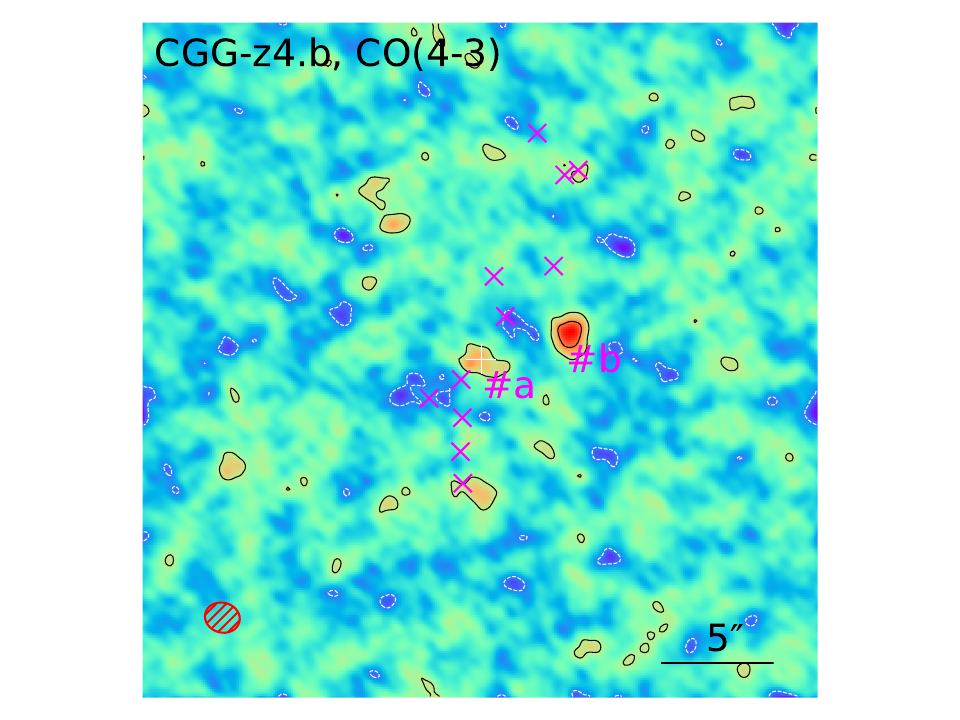}
    \hspace{-1cm}
    \includegraphics[width=0.36\textwidth]{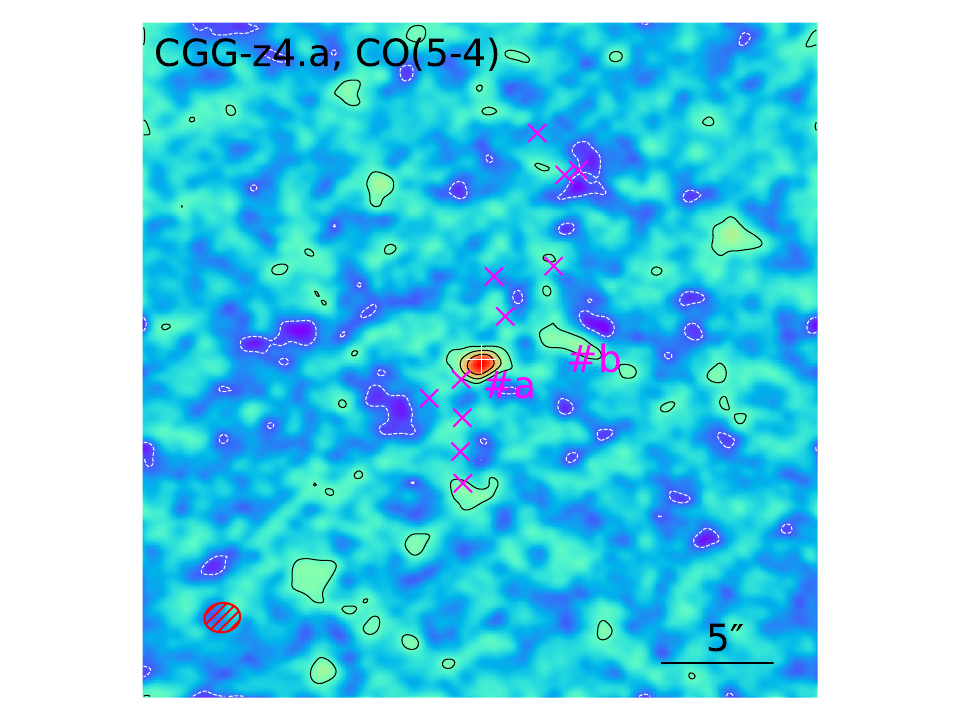}
    \hspace{-1cm}
    \includegraphics[width=0.36\textwidth]{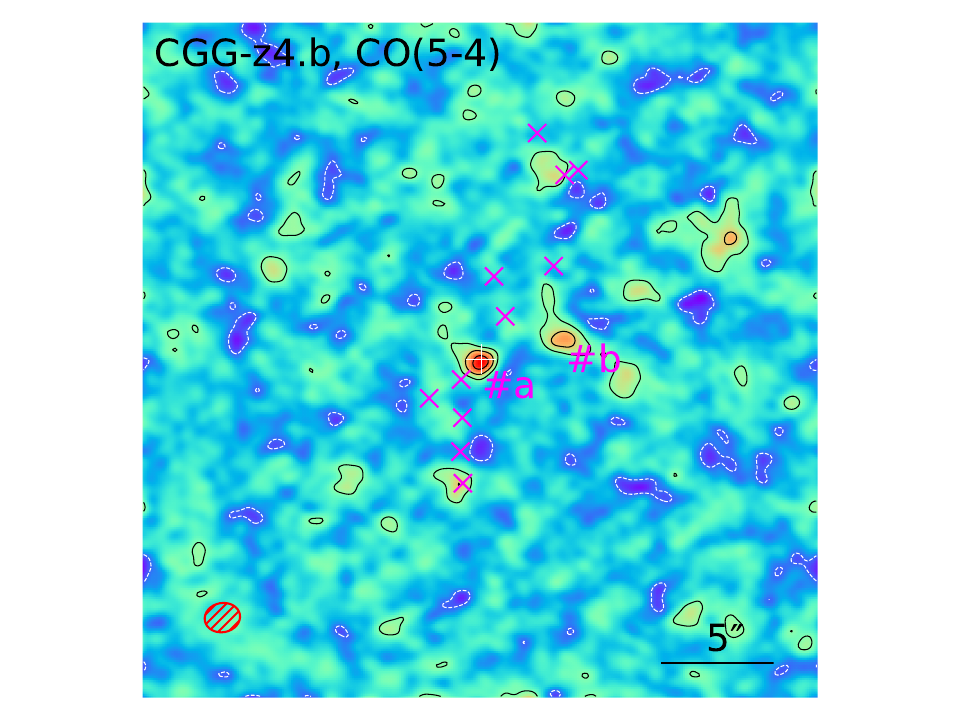}   
    \hspace{-1cm}
    \includegraphics[width=0.36\textwidth]{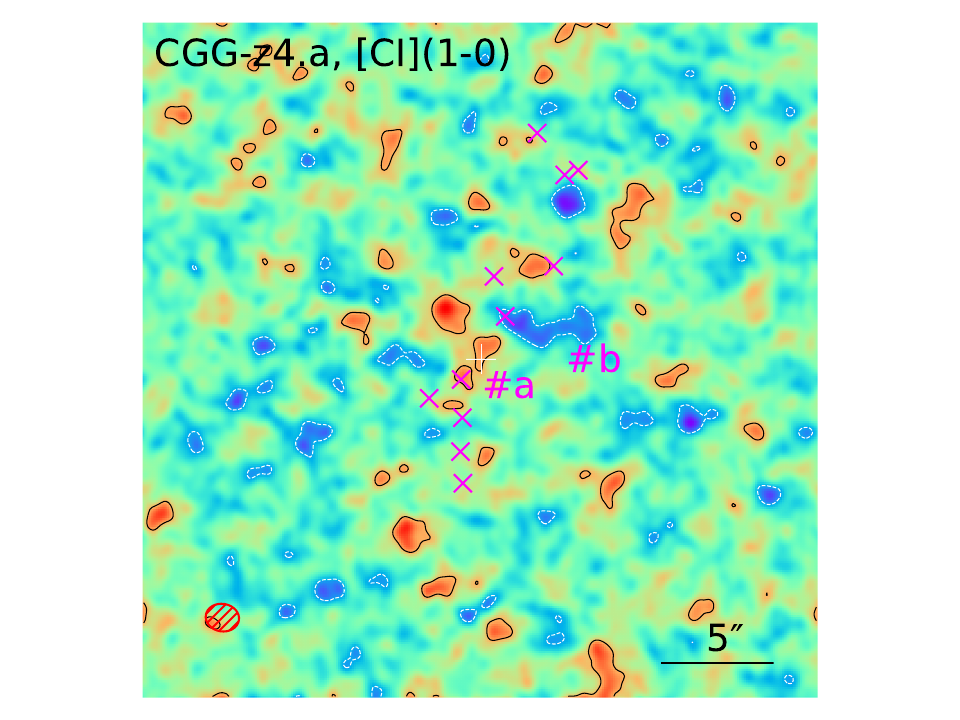}
    \caption{ALMA line and continuum maps, with black lines showing the positive contours and white dashed lines the negative contours, both starting from $2\sigma$ to $6\sigma$ in steps of $2\sigma$; except for the continuum map, where the contours start from $3\sigma$ to $12\sigma$ with $3\sigma$ steps. The size and orientation of the beam are indicated in the lower left corner in red. Magenta crosses mark COSMOS2020 sources.}
    \label{fig:linemaps}
\end{figure*}

\subsection{Selection}

CGG-z4.a was initially selected as a radio source by the VLA 3GHz Large Program in the COSMOS field \citep{Smolcic2017} at $8\sigma$.
It was included in the COSMOS Super-deblended catalog \citep{Jin2018} as an additional radio prior in combination with the low-angular resolution Herschel bands and with no optical/NIR counterpart in the COSMOS2015 catalog \citep{Laigle2016}.
It was observed by an ALMA 3mm line scan as a bright SCUBA-2 source in the \cite{Simpson2019} map, a part of the ALMA project 2021.1.00246.S, (PI: C. Chen).
Our initial FIR SED analysis with deblended FIR photometry suggested a $z_{\rm FIR}\sim6$; this was followed up with an ALMA 3mm line scan in program 2022.1.00884.S (PI: R. Gobat). CGG-z4.b has been observed in the ALMA pointing as well, and is also included among the radio sources in the \cite{Smolcic2017} catalog at $8\sigma$. 
The remaining candidate members of CGG-z4 are selected from the COSMOS2020 CLASSIC catalog \citep{Weaver2022}. We adopted a photometric redshift selection of $4.0<z_{\rm phot}<4.6$ from the EAZY spectral energy distribution (SED) fitting \citep{Brammer2008} with UV/optical to NIR photometry in COSMOS2020, corrected for Galactic extinction using the \citet{Schlafly2011} dust map.  
The galaxies are not part of the area covered by the JWST treasury program COSMOS-Web (ID:1727, PI: J. Kartaltepe, \citealt{Casey2023}).

\subsection{ALMA}
The two ALMA 3mm line scan programs continuously cover the frequency range from 84.1 to 113.2 GHz in Band 3 \citep{ALMAband3}. 
Program 2021.1.00246.S yielded a continuum sensitivity of $0.02\,\rm mJy/Beam$, an angular resolution of $1.6\arcsec$ and a velocity resolution of $84\,\rm km/s$ observed over a $26\,\rm min$ scan.
Program 2022.1.00884.S yielded a continuum sensitivity of $0.02\,\rm mJy/Beam$, an angular resolution of $1.1\arcsec$ and a velocity resolution of $23\,\rm km/s$ observed over three $20\,\rm min$ scans.
ALMA $870\,\rm \mu m$ continuum detections were obtained as part of the program 2016.1.00463.S (PI: Y. Matsuda), in which CGG-z4 was observed over $0.7\,\rm min$ in Band 7 \citep{ALMAband7-2005,ALMAband7-2006} with a continuum sensitivity of $0.27\,\rm mJy/Beam$, an angular resolution of $0.8\arcsec$ and a velocity resolution of $27\,\rm km/s$.
The ALMA data were reduced and calibrated using the standard ALMA CASA pipeline \citep{CASA}. Following our established pipeline from \cite{Jin2019,Jin2022,Jin2024cosbo7}, the calibrated measurement sets were then converted to uvfits format for further analysis in uv space with the GILDAS software \citep{Gildas2013}. 
For the galaxies with a continuum detection, the spectrum was extracted using Gildas \texttt{uv\_fit}, in which we fit $uv$ visibility at each frequency using a point source model fixed at the position of continuum peak. For sources without strong continuum detection, their spectra were extracted based on positions from the COSMOS2020 catalog. We note that program 2021.1.00246.S contains 20 spectral windows (SPWs), and project 2021.1.00246.S contains 12 SPWs. We extracted the 3mm 1D spectrum of each SPW and combined the spectra of all SPWs in the observed frame to enhance the signal-to-noise ratio (S/N) at overlapped frequencies. This resulted an average $1\sigma$ spectral line sensitivity of 0.05 Jy~km~s$^{-1}$~beam$^{-1}$ over a 500 km~s$^{-1}$ width in Band 3.
The continuum fluxes were measured by combining all line-free channels and fitting the sources in $uv$ space with point-source models. The clean residuals indicate the sources are well-fitted and are unresolved in the dust continuum.
The resulting synthesized beam size is $\approx2.7\arcsec$ for the 3mm data, while the $870\,\rm \mu m$ data has a synthesized beam of $\approx0.8\arcsec$.
The spectra of CGG-z4.a and CGG-z4.b are shown in Fig. \ref{fig:color-image}, while line and continuum maps are shown in Fig. \ref{fig:linemaps}.

\subsection{Ancillary data}
We use VISTA images from \cite{Weaver2022} for visualization, as shown in Fig.~\ref{fig:color-image}.
To identify the candidate members in this group, we utilized the optical/NIR photometry and photometric redshifts in the COSMOS2020 catalog \citep{Weaver2022}. We note that the area of CGG-z4 was masked in the COSMOS2020 Farmer catalog, so instead we used the COSMOS2020 Classic catalog which fully cataloged galaxies in this area.
To constrain the FIR SED of CGG-z4, we make use of MIPS \citep{LeFloch2009}, Herschel \citep{Lutz2011PACS,Oliver2012SPIRE}, SCUBA-2 \citep{Simpson2019}, and VLA images \citep{Smolcic2017} to measure the FIR-to-radio photometry with the Super-deblending technique \citep{Jin2018,Liu2018}.

\section{Method}\label{section:method}

\subsection{Line detections and Redshift determination}\label{subsection:line_fitting}

We use the \texttt{LMfit} package \citep{Newville2014} to fit the emission lines and continuum simultaneously. Emission lines were fitted as Gaussians. When multiple emission lines were present in the same spectra, we fixed the sigma value of the Gaussian to be the same when performing the fitting, which, in turn, means the full width at half maximum (FWHM) is fixed. the continuum is fitted as a power law using the formula $y=A\times \nu^{P}$, where A is the amplitude and $P=3.7$ following \cite{Magdis2012,Jin2022}. The line fits are shown in Fig. \ref{fig:color-image} and the derived properties are reported in Table \ref{tab:line}. 
With line detections of CO(4-3) and CO(5-4), we measured a redshift of $z=4.331\pm0.001$ and $4.324\pm0.001$ for CGG-z4.a and b, respectively. The CO(4-3) lines are detected with an integrated line flux significance of 11.6$\sigma$, 8.4$\sigma$, and CO(5-4) is detected with 7.7$\sigma$, 4.8$\sigma$ respectively. The redshift difference between the two sources is $\Delta z=0.007$, which is equivalent to an on-sky projected line of sight distance of $0.829\,\rm pMpc$ ($395\,\rm km/s$).
Notably, we also detect the [C{\sc i}](1-0) emission line in CGG-z4.a with 4.3$\sigma$ and ortho-H$_2$O(1$_{1,0}$-1$_{0,1}$) absorption line in CGG-z4.b  with 3.5$\sigma$. 
The detection of water absorption in CGG-z4.b adds to a growing list of lines observed in the high redshift universe, including the H$_2$O(1$_{1,0}$-1$_{0,1}$) absorption in HFLS3 at $z=6.34$ \citep{Riechers2022} and in ID 9316 at $z=4.07$ \citep{Jin2022}. The water line is an indicator that the galaxy is currently undergoing a starburst, as the intense radiation field from the starburst results in a line excitation temperature that is lower than the CMB temperature, resulting in absorption against the CMB. This also allows the line to be a probe of the CMB temperature \citep{Riechers2022}.

\begin{table*}
\caption[\protect]{Line fitting results for CGG-$z4$ group galaxies.}
\label{tab:line}
\centering
\begin{tabular}{l l l l l l l l}
\hline
\hline
ID & RA, Dec & $z_{\rm spec}$ & I$_{\rm CO(4-3)}$ & I$_{\rm [C_{\rm I}](1-0)}$ & I$_{\rm CO(5-4)}$ & I$_{\rm H_{2}O(1_{1,0}-1_{0,1})}$ &FWHM\\ 
 & [hh:mm:ss, dd:mm:ss](J2000) &  & [Jy km/s] & [Jy km/s] & [Jy km/s] & [Jy km/s] & [km/s] \\
(1) & (2) & (3) & (4) & (5) & (6) & (7) & (8)\\  
\hline
CGG-z4.a & 10:02:40.4313, 01:45:44.0860 & 4.331$\pm$0.001 & 0.67$\pm0.06$ & 0.22$\pm 0.05$ & 1.02$\pm 0.13$ & $<0.11$ ($2\sigma$) & 588\\
CGG-z4.b & 10:02:39.9297, 01:45:46.3900 & 4.324$\pm$0.001 & 0.39$\pm0.05$ & $<0.09$ ($2\sigma$) & 0.55$\pm 0.12$ & -0.15$\pm 0.04$ & 545\\
\hline
\end{tabular}
\tablefoot{Columns: (1) Name; (2) RA, Dec; (3) $z_{\rm spec}$; (4) CO(4-3) line emission; (5) [C{\sc i}](1-0) line emission; (6) CO(5-4) line emission; (7) H$_2$O($1_{1,0}-1_{0,1}$) line absorption; (8) FWHM. Fluxes have been primary beam-corrected.}
\end{table*}

\subsection{Spectral stacking}
To identify potential CO emissions from the optical/NIR-detected members,
we extracted spectra at the positions of all candidate galaxy group members and stacked them at the redshift of CGG-z4.a.  
We calculated the S/Ns at CO(5-4) and CO(4-3) frequencies and we found no definitive detections at the $2\sigma$ level (0.06 Jy km/s for CO(5-4) and 0.04 Jy km/s for CO(4-3)), nor a continuum detection. This indicates that the two optically dark galaxies dominate the total SFR and gas content of this group.

\subsection{SED fitting}\label{subsection:SED}
\begin{figure}
    \centering
    \includegraphics[width=0.50\textwidth]{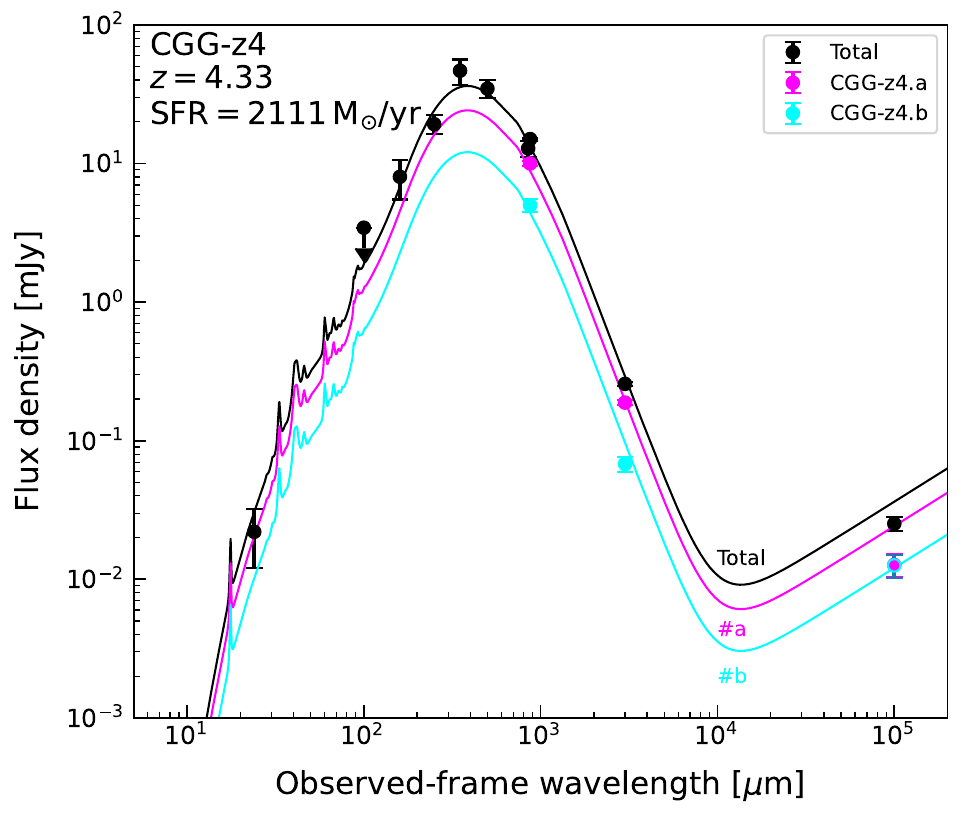}
    \caption{FIR SED of the CGG-z4 integrated group galaxies.
    Photometry data are shown by circles with error bars or downward-pointing arrows for upper limits if $\rm S/N < 3$. The magenta and cyan curves are the total SED scaled by the 870$\,\rm \mu m$ flux of CGG-z4.a and CGG-z4.b, respectively.}
    \label{fig:MICHI2}
\end{figure}

To estimate the physical characteristics of the CGG-$z4$.a and CGG-$z4$.b, we performed a detailed SED analysis by fitting the integrated FIR, (sub-)mm, and radio photometry.
The integrated FIR to radio photometry is measured using the Super-deblending technique \citep{Jin2018,Liu2018}, following the identical pipeline adopted for high-redshift groups as in \cite{Daddi2021,Sillassen2022,Zhou2023,Sillassen2024}.
We used the SED fitting code {\sc MICHI2} \citep{Liu2020,Liu2021} that simultaneously fits components of stellar \citep{Bruzual2003}, mid-IR active galactic nuclei \citep[AGNs,][]{Mullaney2011} and dust \citep{Draine2007}, as well as a radio component derived from IR-radio correlation \citep{Magnelli2015}, without the assumption of energy balance. 
Given the optically dark nature of CGG-z4.a and CGG-z4.a, we fit the MIPS/24$\rm \mu m$, Herschel, SCUBA2, and ALMA photometry, with a mid-IR AGN, warm photodissociation region (PDR), and cold and ambient dust components. 
Figure \ref{fig:MICHI2} shows the full fit, with the photometry used shown in Table \ref{tab:MICHI2}.
For comparison with the ALMA and radio data where the two galaxies are resolved, we scaled the combined SED with the ALMA 870 $\mu$m flux of CGG-z4.a and CGG-z4.b. We observe that scaled SED fits align with the data, with the exception of the radio data of CGG-z4.a, which is below the expected value from the IR-radio correlation. 
When fitting the optically detected COSMOS2020 galaxies, it should be noted that the SED fits from {\sc EAZY} are a linear combination of 12 preselected flexible stellar population synthesis (FSPS) templates. {\sc EAZY}  primarily excels in determining the redshift probability distribution of large galaxy samples. Other SED fitting codes are more suited when estimating the physical properties of individual galaxies. To that end, we use the SED fitting code {\sc Bagpipes} \citep{Carnall2018} to fit the COSMOS2020 classic photometry for physical parameters like stellar mass and star formation rate (SFR). We followed the approach of \cite{Carnall2023,Heintz2023,Jin2023} and assumed a constant star formation history (SFH) and the attenuation curve of \cite{Salim2018}, which mimics a bursty SFH on short timescales. A \cite{Salim2018} curve better represents low-mass, low-metallicity galaxies. We note that we tried using an exponential SFH, which resulted in virtually identical results since our current data only allows us to probe the recent burst for the optical/NIR galaxies due to the outshining effect from the massive young stars in the galaxies. We ran {\sc Bagpipes} with the following parameters: formation mass $\rm log(M_{\rm \star})\in[6.0:12.0]\, log(M_{\rm \odot})$, metallicity $\rm Z\in[0.01:2.00]\, Z_{\rm \odot}$, reddening $\rm A_{\rm v}\in[0.0:4.0],$ and the $U$ parameter, which is the strength of the nebular radiation field to $\log(U) \in[-4.0:-1.0]$. We ran the fitting for each COSMOS2020 galaxy at the spectroscopic redshift of CGG-z4.a ($z=4.331$), assuming they are all part of the galaxy group. 
The resulting SED fits are shown in Fig. \ref{fig:BPSEDS} and a summary of all the values obtained from the SED fitting is shown in Table \ref{tab:SED}.
We note that allowing the photometric redshift to vary in {\sc Bagpipes} results in changes to the posterior fit values that are (at most) at the $1\sigma$ level, with the exception being the age of ID361346 having a higher metallicity of $0.54^{+0.66}_{-0.37}\,\rm Z_{\odot}$ and a younger age of $0.13^{+0.18}_{-0.10}\,\rm Gyr$. This does not change any of the conclusions drawn in this paper. 

\section{Results and analysis}\label{section:results}

\subsection{A $z\approx4.3$ compact galaxy group}
Figure \ref{fig:color-image} presents color images of CGG-z4, comprised of Hyper Suprime-Cam (HSC) and Visible and Infrared Survey Telescope for Astronomy (VISTA) bands with SCUBA2 and ALMA contours overlaid. The Ks-band image highlights the region where the original SCUBA source was detected, while the Ks,J,i true color image shows the ALMA continuum contours of the two optical/NIR dark sources, as well as the optically/NIR, detected galaxies that are within the COMOS2020 catalog. 
The ALMA 3mm spectra for the two sources with multiple line detections, CGG-z4.a and CGG-z4.b, are shown in the bottom panel of Fig. \ref{fig:color-image}. CGG-z4.b is within 1\arcsec of a $z=0.8$ galaxy identified in the COSMOS2020 catalog with ID 357498. Similar to the SMGs of \cite{Smail2021}, CGG-z4.a and CGG-z4.b are not detected in the Ks-band with Ks>25.3 mag in the COSMOS2020 catalog \citep{Weaver2022}, with the combination of no detection blueward of the Ks-band classifying them as Ks-band dropouts. 
Figure \ref{fig:linemaps} Shows the ALMA 3mm continuum map and line emission maps for CGG-z4.a and CGG-z4.b, with the members found in COSMOS2020 overlaid. We see a high sky overdensity of galaxies within a small region, with 13 galaxies over a region of $13\arcsec\times31\arcsec$ ($89\times209\,\rm kpc^2$).  
For any galaxy group or protocluster to grow into a cluster over time, it requires material (other halos with galaxies and gas) to be fed into the structure. It is therefore important to consider the large-scale environment that the group is part of when considering if it will end up as a cluster by $z=0$. Performing an overdensity analysis for galaxies at $z=4.0-4.6$ in the COSMOS field in the same manner as \cite{Brinch2023}, utilizing a weighted adaptive kernel estimator and including the spectroscopic redshifts of Khostovan et al. (in prep.), we find that CGG-z4 appears to be in an overdense environment at the $3\sigma$ level ($\delta_{\rm OD}=2.16$). This significance is similar to other studies that have used SMGs as tracers of overdense environments \citep[see][and references therein]{Calvi2023}. 

\subsection{Stellar mass}
The results of {\sc EAZY} and {\sc Bagpipes} SED fitting are listed in Table \ref{tab:SED}. The EAZY median redshift in the tables is shown to highlight the uncertainty of the redshifts. 
Given the optically dark nature of CGG-z4.a and b, coupled with the fact that they are only detected in IRAC channel 2, we applied two methods to estimate their stellar masses. First, we scaled their IRAC channel 2 peak fluxes to the IRAC fluxes of the COSMOS2020 sources with photometric redshifts between $4.0<z<4.6$ and then scaled the average stellar mass to CGG-z4.a and b. This yielded log$M_*/M_\odot\sim10.4$ for both galaxies. A caveat is that the inferred stellar masses are prone to be uncertain as they are based on one-band photometry. 
Second, given their robust detections in sub-mm and mm wavelengths, we inferred their stellar masses using dust mass with empirical stellar-to-dust mass ratios. By fitting optically thick blackbody models to their FIR and ALMA photometry (e.g., \cite{Lamperti2019}), we obtained dust masses of $\rm log(M_{\rm dust}\,\rm [M_{\odot}])=9.10^{+0.29}_{-0.28}$ and $\rm log(M_{\rm dust}\,\rm [M_{\odot}])=8.80^{+0.29}_{-0.29}$ for a and b, respectively. 
Then we adopted a range of stellar-to-dust mass ratios $M_*/M_{\rm dust}=50-100$, a typical value for starburst galaxies at high-redshifts \citep{Donevski2020}. We obtained stellar masses of $\rm log(M_{\rm \star}\,\rm [M_{\odot}])=10.98^{+0.30}_{-0.37}$ and $\rm log(M_{\rm \star}\,\rm [M_{\odot}])=10.68^{+0.30}_{-0.38}$. The $\rm M_{\rm dust}$-based stellar masses are 0.62 and 0.29 dex higher than IRAC-scaled ones, which is understandable as the IRAC fluxes (rest-frame of 0.8$\,\rm \mu$m) can be severely attenuated in optically dark galaxies (e.g., \citealt{Kokorev2023HSTdark}). To reconcile the results from the two methods, we conservatively adopted the average stellar masses with uncertainties compassing both results, as listed in Table \ref{tab:SED}.
Future observations with JWST will be essential to constrain their stellar content robustly. 

\subsection{Halo mass}\label{sec:halo mass}

We applied five methods to estimate the dark matter halo mass. We followed the identical pipeline by \cite{Sillassen2024}, which was intricately designed for eight massive groups at $1.5<z<4$ in the COSMOS field, where they applied six methods for estimating the dark matter masses, including the stellar-mass-to-halo-mass relations (SHMR), overdensity with galaxy bias, and NFW profile fitting to radial stellar mass densities (see details in \citealt{Sillassen2024}, Sect. 4.4).

First, using the stellar mass of the central galaxy CGG-z4.a with a SHMR from \cite{Behroozi2013}, we obtained a lower limit of halo mass $\rm log(M_{\rm DM}[M_{\rm \odot}])>12.2$. Second, based on the stellar masses above the completeness limit of the COS- MOS survey and an SMF \citep{Muzzin2013}, we obtained a sum of stellar masses within a posterior virial radius of 105 kpc determined on the basis of the best halo mass from different methods. Then we derived a halo mass of $\rm log(M_{\rm DM}[M_{\rm \odot}])=12.4\pm0.4,$ using the dynamical mass-constrained SHMR for $z\sim1$ clusters \citep{van_der_Burg2014}. Third, based on the same total stellar mass, this yielded  $\rm log(M_{\rm DM}[M_{\rm \odot}])=12.6\pm0.3$ by adopting SHMR from \citep{Shuntov2022}. Fourth, based on the overdensity level of CGG-z4 above the whole COSMOS field and adopting a galaxy bias factor, we obtained  $\rm log(M_{\rm DM}[M_{\rm \odot}])=13.5\pm0.3$.
Fifth, utilizing the newly invented NFW profile fitting to the radial stellar mass densities by \cite{Sillassen2024}, we obtained  $\rm log(M_{\rm DM}[M_{\rm \odot}])=13.1_{-0.5}^{+0.2}$. Given the five results above, we adopted their average of $\rm log(M_{\rm DM}[M_{\rm \odot}])=12.8\pm0.4$ as the best estimate.

\subsection{Star formation and ISM properties}

\begin{figure*}
    \centering
    \includegraphics[width=0.49\textwidth]{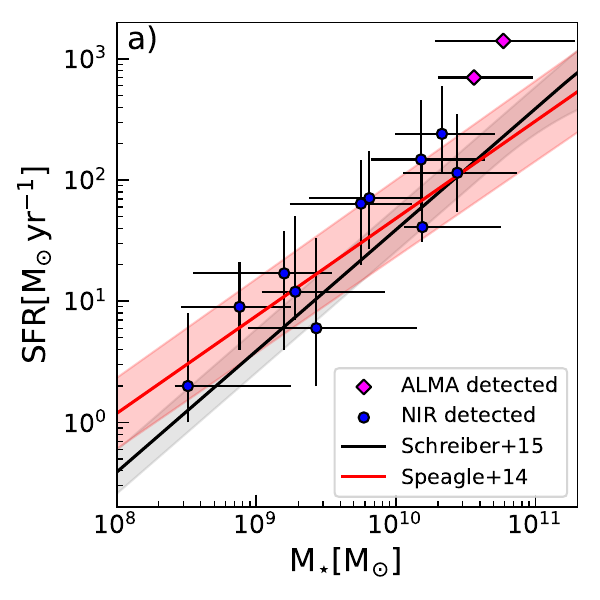}
    \includegraphics[width=0.49\textwidth]{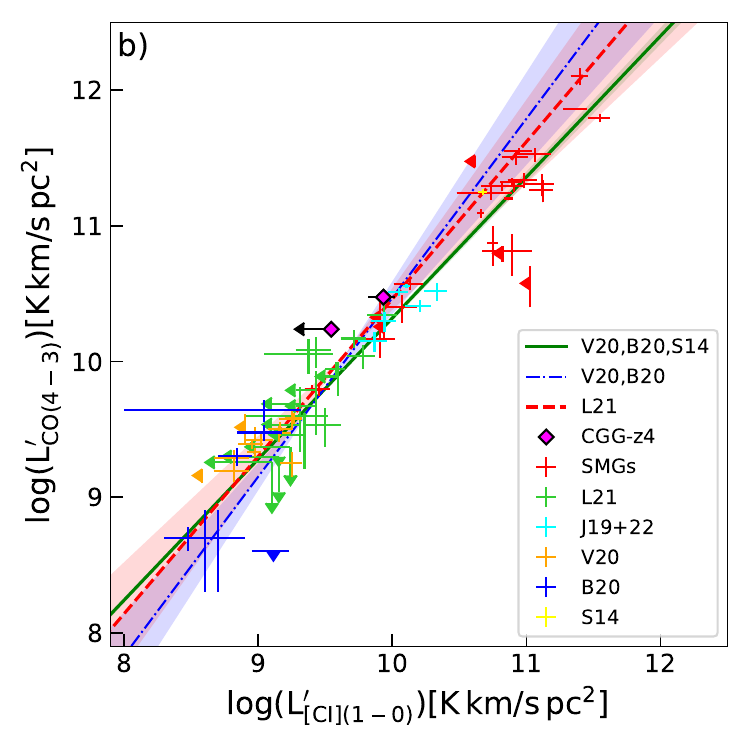}
    \caption{Galaxy main sequence and brightness temperatures for the CGG-$z4$ galaxies. \textbf{(a)} SFR vs. stellar mass. The solid lines indicate the star-forming main sequence (MS) at $z = 4.3$, with the shaded area being the $1\sigma $ scatter \citep{Speagle2014,Schreiber2015}. CGG-z4.a and CGG-z4.b are shown as magenta diamonds, while other galaxies in CGG-z4 are shown as blue circles. \textbf{(b)} The brightness temperature relationship between [C{\sc i}](1-0) and CO(4-3). CGG-z4.a and CGG-z4.b are shown as magenta diamonds. Linear fits are from \cite{Lee2021}, with the data used being $z=2.5$ protocluster galaxies from \cite{Lee2021}(L21). Cold dusty galaxies at $z=3.5-6$ from \cite{Jin2019,Jin2022}(J19+22), a mix of local LIRGs (including AGNs), $z=1$ main sequence galaxies, and DSFGs from \cite{Valentino2020} (V20), main sequence star-forming galaxies at $z\sim1$ from \cite{Boogaard2020} (B20), and the SPT DSFG average from \cite{Spilker2014} (S14). The remaining data comprise SMGs from \cite{Daddi2009,Cox2011,Lestrade2011,Alaghband-Zadeh2013,Aravena2013,Bothwell2013,Omont2013,Canameras2015,Harrington2016,Bothwell2017,Yang2017,Valentino2020}.}
    \label{fig:MS/LL}
\end{figure*}

The best-fit MICHI2 model yields a total IR $\rm SFR=2111\pm98\,\rm M_{\odot}/yr$, mean radiation field of $U=56\pm5$, and dust mass of $\rm log(M_{\rm dust} [M_{\odot}])=9.42\pm0.03$. 
We obtained a total IR luminosity of $\rm log(L_{\rm IR} [L_{\odot}])=13.32\pm0.02$ from integrating the SED fit between 8-1000$\rm \mu m$. 
We find no evidence of mid-IR AGN contribution, with the AGN luminosity being 0.36\% of the total IR luminosity, which is within the uncertainty of the fit. 
To derive the FIR properties of individual galaxies, we assumed that CGG-z4.a and CGG-z4.b share the same SED shape as the integrated SED. We then scaled the SED to the ALMA 870$\,\rm \mu m$ flux of each source, which is $10.0\pm0.4$ mJy and $5.0\pm0.6$ mJy, respectively. The SFR of individual sources is listed in Table \ref{tab:SED}, with IR $\rm SFR=1408\pm86\,\rm M_{\odot}/yr$ and $703\pm65\,\rm M_{\odot}/yr$ for CGG-z4.a and CGG-z4.b, respectively. 
The galaxy main sequence (MS) is shown in Fig. \ref{fig:MS/LL}a, showing both the \cite{Speagle2014} and \cite{Schreiber2015} MS. 
CGG-z4.a and CGG-z4.b are seen as clearly being above the main sequence at $z=4.3$, at $\times6$ and $\times4.5,$ respectively. The most massive member besides CGG-z4.a and CGG-z4.b with optical/NIR data is ID:361388, with a stellar mass of $\rm log(M_{\star}[M_{\odot}])=10.44\pm0.23$ and an SFR of $115^{+121}_{-55}\,\rm M_{\odot}\,\rm yr^{-1}$.  It is placed within the scatter of the \cite{Schreiber2015} main sequence at $z\sim4$. Furthermore, ID:361388 is also the most dust-obscured source in our sample, with $A_{\rm v}=3$. The galaxies of CGG-z4 are a mixture of massive and star-forming galaxies  
with SFRs of $\approx41-241\,\rm M_{\odot}\,\rm yr^{-1}$, along with less massive galaxies with lower SFRs 
with stellar masses as low as $\rm log(M_{\star}[M_{\odot}])=8.51$ and SFRs of $\approx2-17\,\rm M_{\odot}\,\rm yr^{-1}$, as seen in Table \ref{tab:SED}.
The combined stellar mass of the group is $\rm log(M_{\star}[M_{\odot}])=11.29^{+0.38}_{-0.33}$ and the combined SFR is $2837^{+731}_{-472}\,\rm M_{\odot}\,\rm yr^{-1}$.

Figure \ref{fig:MS/LL}b shows the relationship between L$^\prime_{\rm [C{\sc I}](1-0)}$ and L$^\prime_{\rm CO(4-3)}$. When compared to the relations using data from \cite{Spilker2014,Boogaard2020,Valentino2018,Valentino2020,Lee2021}, CGG-z4.a appears to be linear or slightly superlinear, while CGG-z4.b is very faint in [C{\sc i}](1-0) and appears superlinear. The fact that the relationship is superlinear indicates that the CO(4-3) emission is tracing the dense gas that forms stars and as a result is also tracing star formation; meanwhile [C{\sc I}](1-0) traces the total gas mass, resembling the effects of the global Kennicutt–Schmidt law \citep{Schmidt1959,Kennicutt1998,Lee2021}. 

\subsection{Gas mass estimates}
To estimate the gas mass from the dust emission in CGG-z4.a and CGG-z4.b, we used a $850\,\rm \mu m$ conversion from \cite{Dunne2022}. The method was originally pioneered by \cite{Scoville2013,Genzel2015,Scoville2016}. They investigated a number of $\alpha_{\rm X}$ conversion factors by dividing their galaxy sample into two. The first group is extreme starburst "SMGs" containing the high-redshift sub-mm
selected galaxies that were discovered pre-ALMA; as such, these are extreme star-forming systems (based on their detection), along with the local ULIRGs and some LIRGs that have
evidence for being very intensely star-forming and obscured regions where conditions are likely to be extreme \citep{Santos2017, Falstad2021}. 
The second group is main sequence galaxies that contain the
lower luminosity local disc galaxies, along with the LIRGS (which are not
extreme),  intermediate redshift sources selected at 250$\mu$m from
the Herschel-ATLAS, $z=0.35$ galaxies from \cite{Dunne2021}
and the $z<0.3$ VALES galaxies \citep{Hughes2017}, the $z\sim1$
galaxies \citep{Valentino2018, Bourne2019}, and the ASPECs
sources denoted as "MS" in that survey \citep{Boogaard2020}. We refer to Table 1. in \cite{Dunne2022} for a full reference list.

We used an $\alpha_{850}$ value of $7.3\pm0.1\times10^{12}\,\rm W\, Hz^{-1}\, M_{\odot}^{-1}$, such that M$_{\rm gas}=\rm L_{850}/\alpha_{850}$. $\rm L_{850}$ is calculated as follows
\begin{equation}
    \rm L_{850}=4\pi S_{\nu(obs)} \times K \left(\frac{D_{L}^{2}}{1 + z}\right)\, W Hz^{-1}
,\end{equation}
where $\rm D_{L}$ is the luminosity distance, $\rm S_{\nu(obs)}$ is the observed flux
density and K is the K-correction to rest-frame $850\,\rm \mu m$, which is defined as
\begin{equation}
    \rm K =\left(\frac{353\,\rm GHz}{\nu_{\rm rest}}\right)^{3+\beta}\left(\frac{e^{h\nu_{\rm rest}/kT_{d}} - 1}{e^{16.956\,\rm K/T_{d}} - 1}\right)
,\end{equation}
where $\nu_{\rm rest} = \nu_{\rm obs}(1 + z)$, $\rm T_{\rm d}$ is the 
luminosity-weighted dust temperature, from an isothermal
fit to the SED with the dust emissivity, $\beta$.
Given their sub-mm brightness and the fact that they are optical/NIR dark, CGG-z4.a and CGG-z4.b are likely starbursts with optically thick dust emission. To account for this possibility, we performed a modified black body fitting using a thin- and thick-dust model following \cite{Lamperti2019}. The values we found for the thin dust model are: $\rm T_{\rm d}=40.23^{+2.92}_{-2.82}\,\rm K$, $\beta=1.85^{+0.16}_{-0.15}$;  for the thick dust model, these are:  $\rm T_{\rm d}=57.57^{+10.36}_{-11.53}\,\rm K$, $\beta=1.93^{+0.27}_{-0.16}$. 
We note that using a higher $\beta\approx1.9$ for the power law continuum fit changes its value by (at most) $3.4\%$ and does not change the value of any line fluxes at the level of precision reported in this paper.
The resulting K corrections are $\rm K_{\rm thin}=4.8\pm1.3\times10^{-3}$ and $\rm K_{\rm thick}=2.9\pm1.1\times10^{-3}$. By scaling the SCUBA-2 $850\,\rm \mu m$ flux by the ALMA $870\,\rm \mu m$ flux for CGG-z4.a and CGG-z4.b, we find the following luminosities, $\rm L_{850,thin,a}=1.4\pm0.4\times10^{24}\,\rm WHz^{-1}$, $\rm L_{850,thin,b}=7.2\pm2.3\times10^{23}\,\rm WHz^{-1}$, $\rm L_{850,thick,a}=9.0\pm4.0\times10^{23}\,\rm WHz^{-1}$, and $\rm L_{850,thick,b}=4.3\pm1.8\times10^{23}\,\rm WHz^{-1}$. The resulting gas masses are $\rm log(M_{\rm gas,thin,a}[M_{\odot}])=11.29^{+0.12}_{-0.16}$, $\rm log(M_{\rm gas,thin,b}[M_{\odot}])=10.99^{+0.12}_{-0.17}$, $\rm log(M_{\rm gas,thick,a}[M_{\odot}])=11.07^{+0.15}_{-0.24}$, and $\rm log(M_{\rm gas,thick,a}[M_{\odot}])=10.77^{+0.15}_{-0.24}$. In combination with the stellar mass estimates, we can estimate the gas mass fraction $\rm M_{\rm gas}/M_{\star}$, with
CGG-z4.a and CGG-z4.b having inferred gas mass fractions (estimated through Monte Carlo sampling) of $\rm M_{\rm gas}/M_{\star}=1.7^{+3.6}_{-0.9}$ and $\rm M_{\rm gas}/M_{\star}=1.7^{+3.4}_{-0.9}$ using a thin dust model and $\rm M_{\rm gas}/M_{\star}=1.0^{+2.2}_{-0.6}$ and $\rm M_{\rm gas}/M_{\star}=1.0^{+2.1}_{-0.6}$ using a thick dust model.
The ratio M$_{\rm gas}$/SFR approximates the gas depletion timescale, which is the inverse of the star formation efficiency (SFE). Given the gas mass and SFR of CGG-z4.a and CGG-z4.b both have gas depletion times of $140\pm(44,46)\,\rm Myr$ using a thin dust model and $84\pm(36,37)\,\rm Myr$ using a thick dust model. The thick dust model results in 0.22 dex lower gas masses, which, in turn, means that the gas mass factions and gas depletion times are 60\% of the thin dust model values.
For comparison, we used our available CO lines and an $\alpha_{\rm CO}$ conversion to estimate M$_{\rm gas, H2}=\alpha_{\rm CO}\times\rm L^{\prime}_{\rm CO(1-0)}$. Since we did not have data for the CO(1-0) line, we instead converted the $\rm L^{\prime}_{\rm CO(4-3)}$ to $\rm L^{\prime}_{\rm CO(1-0)}$ first. By comparing the flux ratio of CO(5-4) and CO(4-3) lines for CGG-z4.a and CGG-z4.b, we find that they have values of $1.52\pm0.24$ and $1.41\pm0.36$, which is close to the thermalization limit of $\rm (J=5)^{2}/(J=4)^{2}=1.56$  \citep{Narayanan2014}; therefore, we obtained the ratio r$_{\rm L^{\prime}41}=\rm L^{\prime}_{\rm CO(1-0)}/L^{\prime}_{\rm CO(4-3)}\sim1.0$. We note that there is a systematic error in using a value of r$_{\rm L^{\prime}41}=1.0$ that is unaccounted for.   
For the value of $\alpha_{\rm CO}$, we used the canonical value for star-bursting high-$z$ SMGs of $\alpha_{\rm CO}=0.8\,\rm [M_{\odot}\, pc^{-2}\, (K\, km\, s^{-1})^{-1}]$ \citep{Downes1998,Tacconi2008,Carilli2010}. Multiplying by a factor $\times1.36$ to account for the contribution from helium, this gives a gas mass of $\rm log(M_{\rm gas}[M_{\odot}])=10.51^{+0.04}_{-0.04}$ and $\rm log(M_{\rm gas}[M_{\odot}])=10.28^{+0.05}_{-0.06}$ for CGG-z4.a and CGG-z4.b, respectively. These values are $\sim0.7, 0.5$ dex lower than the ones estimated using the fit values and result in both lower gas mass ratio ($0.3^{+0.6}_{-0.1}$ and $0.3^{+0.6}_{-0.2}$) and lower gas depletion times ($23\pm2$ Myr and $27\pm4$ Myr). 

It has been argued in recent times that there is no bimodality in the $\alpha_{\rm CO}$ value between star-bursting SMGs and main sequence galaxies as investigated in \cite{Dunne2022}. 
\cite{Dunne2022} reported a value of $\alpha_{\rm CO}=3.8\pm0.1 \,\rm [M_{\odot}\, pc^{-2}\, (K\, km\, s^{-1})^{-1}]$ (including a factor 1.36 to account for He). \cite{Harrington2021} similarly found $\alpha_{\rm CO}=3.4-4.2 \,\rm [M_{\odot}\, pc^{-2}\, (K\, km\, s^{-1})^{-1}]$ for lensed SMGs at $z\sim1.0-3.5$. Using the $\alpha_{\rm CO}$ value of \cite{Dunne2022}, we obtained estimates for the gas mass: $\rm log(M_{\rm gas}[M_{\odot}])=11.05^{+0.04}_{-0.04}$ and $\rm log(M_{\rm gas}[M_{\odot}])=10.82^{+0.05}_{-0.06}$ for CGG-z4.a and CGG-z4.b. Compared to the two dust models, these estimates are closer to the ones using the thick dust model, though the $1\sigma$ errors overlap with both dust models. The gas mass ratios for CGG-z4.a and CGG-z4.b are $1.0^{+2.1}_{-0.5}$ and $1.2^{+2.3}_{-0.6}$ and gas depletion times are $80\pm9$ Myr and $94\pm15$ Myr, respectively.
Lastly, we can estimate the gas mass from the $[\rm C_{\rm I}](1-0)$ line following \cite{Dunne2022}, giving us    a value of $\alpha_{\rm CI}=16.2 \pm 0.4 \,\rm [M_{\odot}\, pc^{-2}\, (K\, km\, s^{-1})^{-1}]$ for SMGs. We used the $2\sigma$ upper limit for CGG-z4.b. We estimated the gas masses for CGG-z4.a and CGG-z4.b as $\rm log(M_{\rm gas}[M_{\odot}])=11.14^{+0.09}_{-0.11}$ and $\rm log(M_{\rm gas}[M_{\odot}])<10.66$. We note that using Eq. 8 in \cite{Dunne2022} also gives consistent results for the gas mass.
This results in gas mass ratios of $1.2^{+2.5}_{-0.6}$ and $<2.3$ and gas depletion times of $99\pm23$ Myr and $<69$ Myr, respectively. The estimate for CGG-z4.a is in agreement with the ones using the two dust models (with the thick dust model being closer) and an $\alpha_{\rm CO}=3.8 \,\rm [M_{\odot}\, pc^{-2}\, (K\, km\, s^{-1})^{-1}]$; whereas for CGG-z4.b, the upper limit is in a good general agreement with all the other estimates.

A summary of the different estimates 
is given in Table \ref{tab:Gasvalue}.
Apart from the upper limit based on using $[\rm C_{\rm I}](1-0)$, the ratio between the objects is consistent, with the gas mass ratio and gas depletion times essentially being the same. To constrain the estimates of M$_{\rm gas}$ better, observations of the $115\,\rm GHz$ CO(1-0) line are needed.
\section{Discussion}\label{section:discussion}

\subsection{Line ratios and ISM conditions}
\begin{figure}
    \centering
    \includegraphics[width=0.5\textwidth]{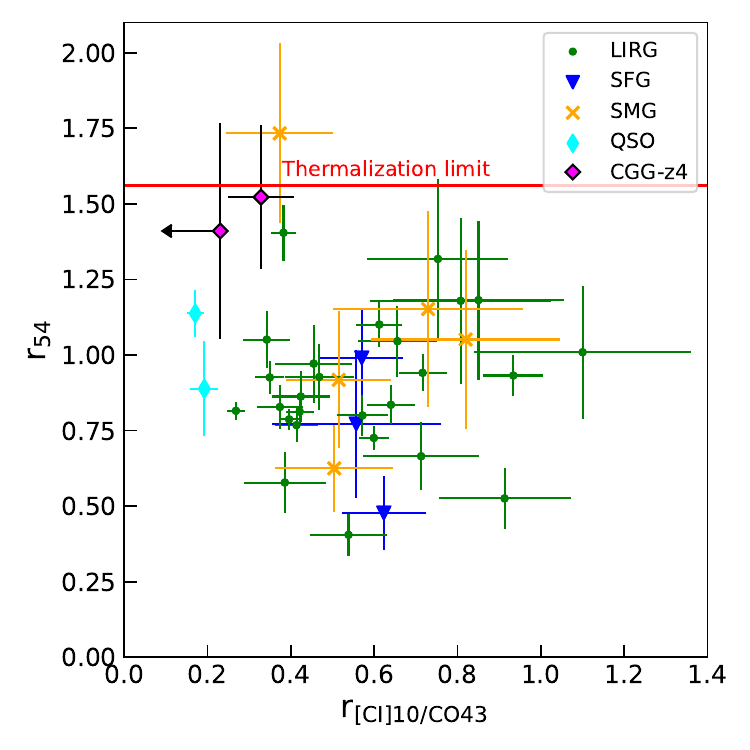}
    \caption{Flux ratios of r$_{\rm [CI]10/CO43}$
    and r$_{54}$. The thermalization limit for r$_{54}$ is given by $(\rm J_{\rm up}=5)^2/(\rm J_{\rm up}=4)^2=1.56$ and is shown as a red line. CGG-z4.a and CGG-z4.b are shown as magenta diamonds. Data is LIRGs from \cite{Lu2017}, SFGs from \cite{Valentino2020}, SMGs from \cite{Bothwell2017,Valentino2020}, and QSOs from \cite{Barvainis1997,Pety2004}.}
    \label{fig:lineratios}
\end{figure}

To investigate the physical state of CGG-z4.a and CGG-z4.b, we compare their available line ratios with public data from different types of sources. The literature sources include luminous IR galaxies (LIRGs) from \cite{Lu2017}, star-forming galaxies (SFGs) from \cite{Valentino2020}, SMGs from \cite{Bothwell2017,Valentino2020}, QSOs from \cite{Barvainis1997,Pety2004}. 
Figure \ref{fig:lineratios} shows the relationship between the flux ratios of 
r$_{\rm [CI]10/CO43}=[\rm C_{\rm I}](1-0)/CO(4-3)$ and r$_{54}=\rm CO(5-4)/\rm CO(4-3)$. The ratio r$_{\rm [CI]10/CO43}$ is a proxy for the ratio between the amount of total gas present in the galaxy and the amount of dense gas present to form stars, which, in turn, means it is indicative of the ratio between the total gas mass and the SFR. The low values of r$_{\rm CI43}$ for CGG-z4.a,b are consistent with our low estimates for the gas depletion time of the galaxies. The values of r$_{\rm CI43}$ are close to the ones that are lowest in our reference sample; namely, the two quasars Cloverleaf \citep{Barvainis1997} and PSS-2322+1944 \citep{Pety2004}. However, in the case of these quasars, 
the detectability of molecular and neutral atomic transitions is in large part due to the high emissivity of the gas and magnification from gravitational lensing, as opposed to a high (dense) gas mass.
The r$_{54}$ ratio describes the excitation state of the galaxies. However, we note that to fully characterize the galaxies, a full CO SLED should be constructed. CGG-z4.a and CGG-z4.b have relatively high values of r$_{54}$, close to the 1.56 limit for a fully thermalized gas.
The high values of r$_{54}$ can indicate different gas conditions in the two galaxies, such as high temperatures, high densities, and/or high pressures. 

We further attempted to investigate the gas density using a PDR model. The PDR model combines available line ratios to constrain the radiation field intensity and the hydrogen nucleus volume density. To constrain the PDR model, typically three line ratios are needed. Utilizing the Python package photodissociation region toolbox \citep[pdrtpy,][]{Pound2023}, we found the Wolfire-Kaufman 2006 model \citep{Kaufman2006}, with a constant density and solar metallicity, to have three available line ratios: r$_{\rm [CI]10/CO43}$, r$_{\rm [CI]10/CO54}$, and r$_{\rm [CI]10/I_{\rm FIR}}$. Here, $\rm I_{\rm FIR}$ is the total FIR intensity, which we have access to from the SED fitting, and (as with the SFR) we scaled to CGG-z4.a and CGG-z4.b using their ALMA $870\,\rm \mu m$ flux. 
We determine a value from the full SED of $\rm I_{\rm FIR,8-1000\, \mu m}=(4.35\pm0.20)\times10^{-16}\,\rm W m^{-2}$.
The package was run using the built-in emcee Markov chain Monto Carlo sampler \citep{Foreman-Mackey2013}, with 2000 steps and 200 walkers to sample the posterior distribution for the model parameters.
For CGG-z4.a, the density is $\rm n_{\rm H}=(3.14\pm 0.67)\times10^{4}\,\rm cm^{-3}$ and radiation field $\rm G_{0}=(1.17\pm 0.49)\times10^{3}$ Habing (the local Galactic interstellar radiation field, which has a value of $1.6\times10^{-6}\,\rm W m^{-2}$). For CGG-z4.b, we only have an upper limit available for the [CI](1-0) line and therefore the pdr models are poorly constrained. Given the available models, we can determine that $\rm n_{\rm H}>3.0\times10^{4}\,\rm cm^{-3}$ and $\rm G_{0}>10.0$ Habing.

To make a comparison with SMGs in the literature, we used a wavelength range of $42.5-122.5\,\rm \mu m$ in the rest-frame \citep{Wardlow2017} and determined $\rm I_{\rm FIR,42.5-122.5\, \mu m}=(2.61\pm0.12)\times10^{-16}\,\rm W m^{-2}$. We obtained $\rm n_{\rm H}=(3.08\pm0.67)\times10^{4}\,\rm cm^{-3}$, $\rm G_{0}=(6.63\pm2.62)\times10^{2}$ Habing for CGG-z4.a and $\rm n_{\rm H}>3.0\times10^{4}\,\rm cm^{-3}$ and $\rm G_{0}>6.0$ Habing for CGG-z4.b.
Compared to the SMGs in \cite{Wardlow2017}, which have densities of $\rm n_{\rm H}\sim3\times10^{4}\,\rm cm^{-3}$ and intensities of $\rm G_{0}\sim7\times10^{3}$ Habing, CGG-z4.a has a similar density and an intensity that is slightly below their sample.
It should be noted that the \cite{Wardlow2017} sample mainly consists of $z=1-3$ galaxies. 
Their lensed galaxy G15v2.779 at $z=4.243$ has the closest redshift to CGG-z4.a, which shows a similar $\rm I_{\rm FIR,42.5-122.5\, \mu m}$ of $2.93\times10^{-16}\,\rm W m^{-2}$ after a correction for magnification. 

\subsection{Gas depletion times}
\begin{figure}
    \centering
    \includegraphics[width=0.5\textwidth]{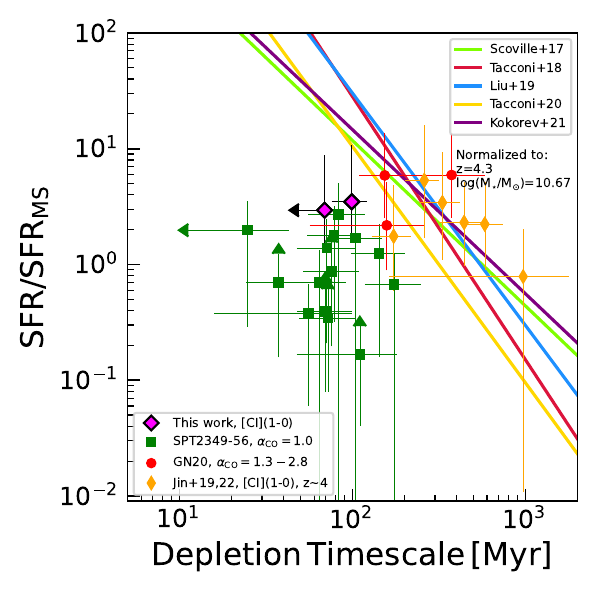}
    \caption{Offset to the \cite{Speagle2014} galaxy main sequence vs. gas depletion time in Myr. The solid lines indicate the scaling relations of \cite{Scoville2017,Tacconi2018,Liu2019,Tacconi2020,Kokorev2021} normalized to $z=4.3$ and $\rm log(M_{\star}[M_{\odot}])=10.67$. CGG-z4.a and CGG-z4.b are shown as magenta diamonds, SPT2349-56 galaxies are shown as green squares \citep{Miller2018,Hill2020,Hill2022}, GN20 galaxies are shown as red circles \citep{Tan2014}, and the \cite{Jin2019,Jin2022} galaxies are shown as orange diamonds. Depletion timescales were calculated for the SPT2349-56 galaxies using a gas mass with an $\alpha_{\rm CO}=1.0$ and a CO(4-3)/CO(1-0) line ratio $\rm r_{41}=0.6\pm0.05$ \citep{Spilker2014}. The gas masses of each GN20 galaxy were calculated using the average of $\alpha_{\rm CO}$ derived from the gas-to-dust mass ratio method and dynamical modeling. The gas masses for the galaxies of \cite{Jin2019,Jin2022} were estimated from [CI](1-0) line data in each paper.
    }
    \label{fig:tdeplvsDMS}
\end{figure}

To compare the efficiency with which these galaxies form stars to test if CGG-z4.a and CCG-z4.b differ from both other SMGs and field galaxies, we compared the depletion times of the CCG-z4 galaxies with those from SPT2349-56 and the sample of $z>4$ SMGs of \cite{Jin2019,Jin2022}. We compared these to the scaling relations of field samples \cite{Scoville2017,Tacconi2018,Liu2019,Kokorev2021}, normalized to a redshift of $z=4.3$ and 
$\rm log(M_{\star}[M_{\odot}])=10.67$ (the average stellar mass of CGG-z4.a and CGG-z4.b). 
We use the gas depletion times estimated using the [CI](1-0) line for CGG-z4.a and CGG-z4.b, since it is the estimate least reliant on conversions that could have systematic uncertainties, such as the choice of dust model or the conversion from CO(4-3) to CO(1-0).
From Fig. \ref{fig:tdeplvsDMS}, we see that the CGG-z4.a and CGG-z4.b both have low depletion times (high SFEs) and are above the main sequence, being slightly below the field relation at $z=4.3$. CGG-z4.b appears to be more offset from the field relations than CGG-z4.a, with both being closest to the relations from \cite{Scoville2017} and \cite{Tacconi2020}. When comparing with the galaxies of \cite{Jin2019,Jin2022}, they appear to be following the field relation, being similarly offset from the galaxy main sequence as CGG-z4.a and CGG-z4.b and having higher gas depletion times. 
The GN20 protocluster galaxies follow the field relation and are close to the ones from \cite{Jin2019,Jin2022}; however, it is worth noting that their gas masses were calculated using an $\alpha_{\rm CO}$ derived from a gas-to-dust mass ratio method and dynamical modeling, which differs for each object.
Compared to the galaxies of SPT2349-56, both CGG-z4.a and CGG-z4.b have similar depletion times, but the galaxies of SPT2349-56 are less offset from the main sequence; therefore, they appear offset from the field relations. This offset to the field relation can be explained by multiple factors. First, the gas mass used to calculate the depletion times was estimated using a $\alpha_{\rm CO}$ numerical value of 1.0 and taken from \cite{Hill2020}, where they argued in favor of the use of this value for the sake of simplicity. Since they have a mix of the main sequence and starburst galaxies, it is possible that some of them should instead be using a $\alpha_{\rm CO}$ numerical value of $\sim4$ (or even higher) due to the low metallicity of main sequence galaxies at high-z (i.e., the evolution of mass-metallicity relation and the dependence of $\alpha_{\rm CO}$ versus metallicity),
which would place the galaxies closer to the field relations. 
Secondly, it is possible that the stellar masses of \cite{Hill2020} could have a large uncertainty due to such factors as the unknown IMF and metallicity. Therefore, these values could be overestimated, resulting in an overly low offset from the \cite{Speagle2014} main sequence.  

\subsection{Onset of quenching}
With the low gas depletion times (high SFE), on the order of $\sim100\,\rm Myr$ for the two dusty galaxies in CCG-z4, it is possible that they will become quiescent at $z\sim4.0$. Under the assumption that their SFE is constant, all the gas is converted into stars, and there is no gas replenishment, their final stellar masses would be $\sim10^{11}\,\rm M_{\odot}$ (with some variation, depending on the gas estimate used). 
This stellar mass is consistent with that of the massive quiescent galaxies in $z>3$ dense environments \citep{Kubo2021quiescent,Kubo2022quiescent,McConachie2022,Jin2024}.
For comparison, the large-scale structure Cosmic Vine \citep{Jin2024} at $z=3.44$ contains two massive quiescent galaxies with stellar masses close to $10^{11}\,\rm M_{\odot}$, and with star-formation histories from SED fitting suggesting they were quenched at $4 < z < 6$. The fact that one of these massive galaxies has formed outside the protocluster core region disfavors environmental quenching with the prerequisite of a virialized cluster core.
\cite{Jin2024} argued that it is likely the two galaxies were quenched by merger-triggered starbursts in the past 500 Myr. In comparison, CGG-z4.a and CGG-z4.b are still in their starbursting phase at $z=4.33$. The time between $z=4.33$ and $z=4.0$ is $\approx 140\,\rm Myr$, and given their short gas depletion times it is possible for the CGG-z4 to evolve into galaxies similar to those of the Cosmic Vine. Due to the CGG-z4 galaxies' optical/NIR dark nature, it is difficult to ascertain the morphology of the galaxies and comment on whether the galaxies have undergone a merger, similar to what is observed with galaxy A in the Cosmic Vine, where a compact bulge and tidal tail has been observed, both servings as indicators of a merger. Studying the morphology of the CGG-z4 optical/NIR galaxies would require deeper photometry than what is currently available in this part of the COSMOS field.

\subsection{Comparison with literature samples}
Here, we consider how CGG-z4 compares to other compact galaxy groups in the literature. 
As a compact galaxy group at $z\sim5.2$ \citep{Jin2023}, CGG-z5 has six members within a projected area of $1.5\arcsec\times3\arcsec$ (10$\times$20 kpc$^2$), 
which is more compact compared to the 13 members of CGG-z4 that are over a projected area of $13\arcsec\times31\arcsec$ ($89\times209\,\rm kpc^2$). The galaxies of CGG-z5 have low stellar masses and SFRs, which is in line with the low mass and low SFR galaxies in CGG-z4. The galaxies of CGG-z5 also appear to have lower A$_{\rm V}$'s ($\approx0.38$), higher metallicities ($\approx0.54\,\rm Z_{\odot}$), and slightly higher ages ($\approx0.14\,\rm Gyr$) to those in CGG-z4. CGG-z5 is expected to merge and grow into a $\rm log(M/M_{\rm \odot})\sim10.7$ galaxy at $z\sim4$ with a merging timescale of $\sim400\,\rm Myr$ when comparing with similar galaxies to CGG-z5 in the {\sc EAGLE} simulation \citep{Crain2015,Schaye2015}. 

The compact group HPC1001 at $z=3.61$ \citep{Sillassen2022} has a similar total mass and $\sim1/3$ the SFR to CGG-z4, but a higher number density, with ten total members and eight members within $10\arcsec\times10\arcsec$ ($70\times70\,\rm kpc^2$). Comparing the most massive galaxy in the group, HPC1001.b (with a mass of $10^{11}\,\rm M_{\odot}$) to CGG-z4.a and CGG-z4.b, it is placed $\times3$ below the main sequence at $z=3.7$; whereas the two galaxies in CGG-z4 are $\times6$ and $\times4.5$ above the main sequence, respectively, at $z=4.3$. Utilizing a dust-to-gas mass ratio of 1:100, HPC1001.b has a gas mass ratio of $\rm M_{gas}/M_{\star}=0.2\pm0.1$. Nonetheless, their gas depletion timescales are comparable, $200\,\rm Myr$ for HPC1001.b and $\approx100\,\rm Myr$ for the CGG-z4 galaxies, with some variation depending on the method used, with the exception being the method using a $\alpha_{\rm CO}$ numerical value of 0.8.  

A group that appears to have similar galaxies to CGG-z4 is the star-bursting galaxy group AzTEC5 at $z\sim3.6$ containing four members within $5\arcsec\times5\arcsec$ \citep[$35\times35\,\rm kpc^2$,][]{Guijarro2018}. AzTEC5-1,2 have ALMA 870$\,\rm \mu m$ counterparts with SFRs and dust masses on a similar level to CGG-z4.a,b, with the ratio between the two galaxy pairs even being similar. Given that the stellar mass of CGG-z4.b is likely overestimated, the ratios of the stellar mass may also be similar; however, in the case of AzTEC5, it is the less actively star-forming galaxy AzTEC5-1 that has the higher stellar mass.

A protocluster found close to the redshift of CCG-z4 is GN20 at $z=4.05$ \citep{Tan2014}. The structure contains the GN20 SMG, which is one of the brightest sources in the GOODS-N field, as well as two additional SMGs, GN20.2a and GN20.2b, located within $\sim25\arcsec$(projected $170\,\rm kpc$) of GN20. Much like CCG-z4, the GN20 galaxies are all starburst with large amounts of molecular gas surrounded by 14 B-band dropout galaxies within $25\arcsec$ of GN20.  
Their gas depletion times are similar to CGG-z4 (at least when comparing their CO estimates of $\sim200\,\rm Myr$). While the CGG-z4 galaxies are optical/NIR dark, the GN20 galaxies do have counterparts, although they appear offset from the sub-mm emission.  

We also compare CGG-z4 to the $z=4.3$ protocluster SPT2349-56. The protocluster has a 130 kpc core region that contains 14 star-forming (total $\rm SFR \sim 5000\, M_{\odot} yr^{-1}$) and massive galaxies ($\rm M_{\star} \sim 10^{10}\, M_{\odot}$)
\citep{Miller2018,Hill2020,Hill2022}. The size of the structure and number of sources in the core are comparable to CGG-z4, but the SPT sources are all highly star-forming and detected in either CO(4-3) or [CII] emission. The gas masses of CGG-z4.a and CGG-z4.b are slightly higher compared to the SPT galaxies, with ranges between $\rm 9.30<log(M_{\rm gas}[M_{\odot}])<10.88$, while the stellar masses are comparable with ranges between $\rm 10.08<log(M_{\rm gas}[M_{\odot}])<11.35$. While the gas mass ratios are generally lower for the SPT galaxies, ranging from $0.06-0.77$, their gas depletion times are similar to the CGG-z4 galaxies, ranging from $\rm 43\, Myr-119\, Myr$.

\subsection{Comparison with simulations}
To understand how CGG-z4 would evolve with time, we compare it to massive structures in simulations with similar dark matter halo mass, $\rm log(M_{\rm DM}[M_{\rm \odot}])=12.8\pm0.4$. Comparing with the curves of \cite{Chiang2013} for the most massive progenitor of present-day clusters in a cluster halo merger tree, we find that CGG-z4 is expected to grow into a Virgo- or Coma-like cluster over the next 10 billion years. It should be noted that halo mass estimates can vary greatly between different models (compare the different estimates in Sect. \ref{sec:halo mass}) due to the different assumptions that go into estimating the dark matter halo mass. Given these caveats, a conservative prediction would be that CGG-z4 is likely to grow into a $\rm log(M_{\rm DM}[M_{\rm \odot}])>14.0$ by $z=0$. 
The presence of highly star-forming galaxies, with large amounts of dense gas in a compact environment, aligns with what we expect from a protocluster in its growing phase at $z>3$ \citep{Shimakawa2018}.
In the case that CGG-z4 will go on to form a galaxy cluster, we attempt to assess the fate of the two most massive galaxies, CGG-z4.a and CGG-z4.b, by making a comparison with the TNG300 simulation \citep{Pillepich2018TNG300}. 
\cite{Taborda2023} selected 280 systems with $M_{200}>10^{14}\,\rm M_\odot$ at $z=0$ in the TNG300 simulation and traced their progenitors and proto-BCGs at high redshift.
Using the results of BCG progenitors by \cite{Taborda2023}, the BCG stellar masses at $z=4.3$ are expected to be on the order of $\rm log(M_{\rm BCG}[M_{\odot}])=10.5^{+0.7}_{-0.9}$,   consistent with the stellar masses of CGG-z4.a and CGG-z4.b. 
On the other hand, the halo mass of CGG-z4 $\rm log(M_{DM}[M_{\odot}])=12.8\pm0.4$ is above the statistical mass $\rm log(M_{200}[M_{\odot}])=12.3^{+0.3}_{-0.5}$ of the clusters at $z\sim4$. These comparisons support the idea that CGG-z4.a and CGG-z4.b are BCGs undergoing formation in a massive cluster.

\section{Conclusions}
Thanks to the use of ALMA and ancillary data in the COSMOS field, we have discovered a compact galaxy group, CGG-z4, hosting two optically/NIR dark galaxies at z=4.3. We report the following conclusions:
   \begin{enumerate}
      \item CCG-z4 contains two optically/NIR dark galaxies with spectroscopic redshifts at $z=4.3$ and 11 optical/NIR-detected candidate members, with photometric redshifts at $4.0<z<4.6$. The galaxies are spread out over a projected area of $13\arcsec\times31\arcsec$ ($89\times209\,\rm kpc^2$).
      
      \item The two optically/NIR dark galaxies, CGG-z4.a and CGG-z4.b, both have robust detections of CO(5-4) and CO(4-3) emission from ALMA 3mm line scans. CGG-z4.a has a detection of [CI](1-0) emission, while CGG-z4.b has a detection of H$_2$O($1_{1,0}-1_{0,1}$) in absorption. We performed spectral stacking of all the optically/NIR-detected galaxies in CGG-z4 and found no definitive detection of spectral lines at the 2$\sigma$ level.
      
      \item Using dust continuum, CO, and CI lines as gas tracers, we found CGG-z4.a and CGG-z4.b are starburst galaxies with large amount of gas reservoirs (log($\rm M_{\rm gas}[M_{\odot}]$)$\sim11.0$), massive stellar masses (log($\rm M_{\rm \star}[M_{\odot}]$)$\sim10.7$), and short gas depletion times ($\sim100\,\rm Myr$). 
      
      \item We compared the line ratios of CGG-z4.a and CGG-z4.b with literature sample and found the ratio $\rm r_{54}$ is close to the thermalization, while $\rm r_{CI43}$ is among the lowest values compared to galaxies from the literature, indicating a high excitation of the ISM and efficient star formation.
      
      \item The high SFEs in CGG-z4.a and CGG-z4.b suggest the onset of quenching. Assuming all of the gas reservoirs are converted to stars under the current SFR, these galaxies would grow $\sim2\times$ their stellar masses up to $\sim10^{11}\,\rm M_\odot$ and would enter a quiescent phase at $z\sim4.0$, which would be similar to massive quiescent galaxies found in other overdense structures at $z>3$.
      
      \item We used multiple approaches to estimate the dark matter halo mass of CGG-z4 and found a best estimate of log($\rm M_{\rm DM}[M_{\odot}]$)$=12.8\pm0.4$. 
      Comparisons with the simulations suggest that CGG-z4 is a forming protocluster, which is likely to form a cluster with $\rm M>10^{14}\, M_\odot$ by $z\sim0$, and the two massive dusty galaxies will form BCGs in the cluster.  
   \end{enumerate}
Spectroscopic follow-up of the nearby optically/NIR detected galaxies in CGG-z4 would further improve the completeness of the membership. JWST follow-up would also robustly constrain the stellar mass of optically/NIR dark galaxies, and further ALMA [CII] observations would be the key to unveiling the inter-group medium and allow us to assess the dynamic state of this structure. 

\begin{acknowledgements}
The Cosmic Dawn Center (DAWN) is funded by the Danish National Research Foundation under grant No. 140. SJ and ML acknowledge the financial support from the European Union’s Horizon Europe research and innovation program under the Marie Skłodowska-Curie grant agreement No. 101060888 and  101107795. JH acknowledges support from the ERC Consolidator Grant 101088676 (VOYAJ).
This paper makes use of the following ALMA data: ADS/JAO.ALMA\#2016.1.00463.S, ADS/JAO.ALMA\#2021.1.00246.S, ADS/JAO.ALMA\#2022.1.00884.S. ALMA is a partnership of ESO (representing its member states), NSF (USA), and NINS (Japan), together with NRC (Canada), NSTC and ASIAA (Taiwan), and KASI (Republic of Korea), in cooperation with the Republic of Chile. The Joint ALMA Observatory is operated by ESO, AUI/NRAO and NAOJ.
We acknowledge the following open-source
software packages used in the analysis: Astropy \citep{AstropyCollaboration2013}, Matplotlib \citep{Hunter2007}, Numpy \cite{Harris2020}, Scipy \citep{Virtanen2020}, Uncertainties: a Python package for calculations with uncertainties, Eric O. Lebigot, \url{http://pythonhosted.org/uncertainties/}.
\end{acknowledgements}

\bibliographystyle{aa_url}
\bibliography{bib}
\begin{appendix} %First appendix
\onecolumn
\section{Photometry used for combined SED fitting of CGG-z4.a+b}
\begin{table}[h]
\caption[\protect]{Data used for SED fitting of CGG-$z4$.a,b.
}
\label{tab:MICHI2}
\centering
\renewcommand*{\arraystretch}{1.5}
\begin{adjustbox}{max width=\textwidth}
\begin{tabular}{l l l l l l l l l l l}
\hline
\hline
ID & S$_{24\rm \mu m}$ & S$_{100\rm \mu m}$ & S$_{160\rm \mu m}$ & S$_{250\rm \mu m}$ & S$_{350\rm \mu m}$ & S$_{500\rm \mu m}$ & S$_{850\rm \mu m}$ & S$_{872\rm \mu m}$ & S$_{3\rm mm}$ & S$_{10\rm cm}$\\ 
 & [mJy] & [mJy] & [mJy] & [mJy] & [mJy] & [mJy] & [mJy] & [mJy] & [mJy] & [mJy] \\
(1) & (2) & (3) & (4) & (5) & (6) & (7) & (8) & (9) & (10) & (11)\\  
\hline
CGG-z4.a+b & 0.022$\pm$0.010 & <3.44 & 8.02$\pm$2.52 & 19.22$\pm$3.01 & 46.71$\pm$9.60 & 34.68$\pm$5.14 & 12.86$\pm$1.77 & 15.01$\pm$0.35 &  0.257$\pm$0.009 & 0.025$\pm$0.003\\
\hline
\end{tabular}
\end{adjustbox}
\tablefoot{CGG-z4.a,b were fitted using the SED fitting code of \cite{Liu2020,Liu2021}. Columns: (1) Name; (2) Mips 24 $\rm\mu m$; (3) PACS 100 $\rm\mu m$; (4) PACS 160 $\rm\mu m$; (5) SPIRE 250 $\rm\mu m$; (6) SPIRE 350 $\rm\mu m$; (7) SPIRE 500 $\rm\mu m$; (8) SCUBA2 850 $\rm\mu m$; (9) ALMA 872 $\rm\mu m$; (10) ALMA 3 mm; (11) VLA 10 cm.}
\end{table} 

%\clearpage
\section{Bagpipes SED fits}
\begin{figure}[h]
    \centering
    \begin{overpic}[width=0.48\linewidth]{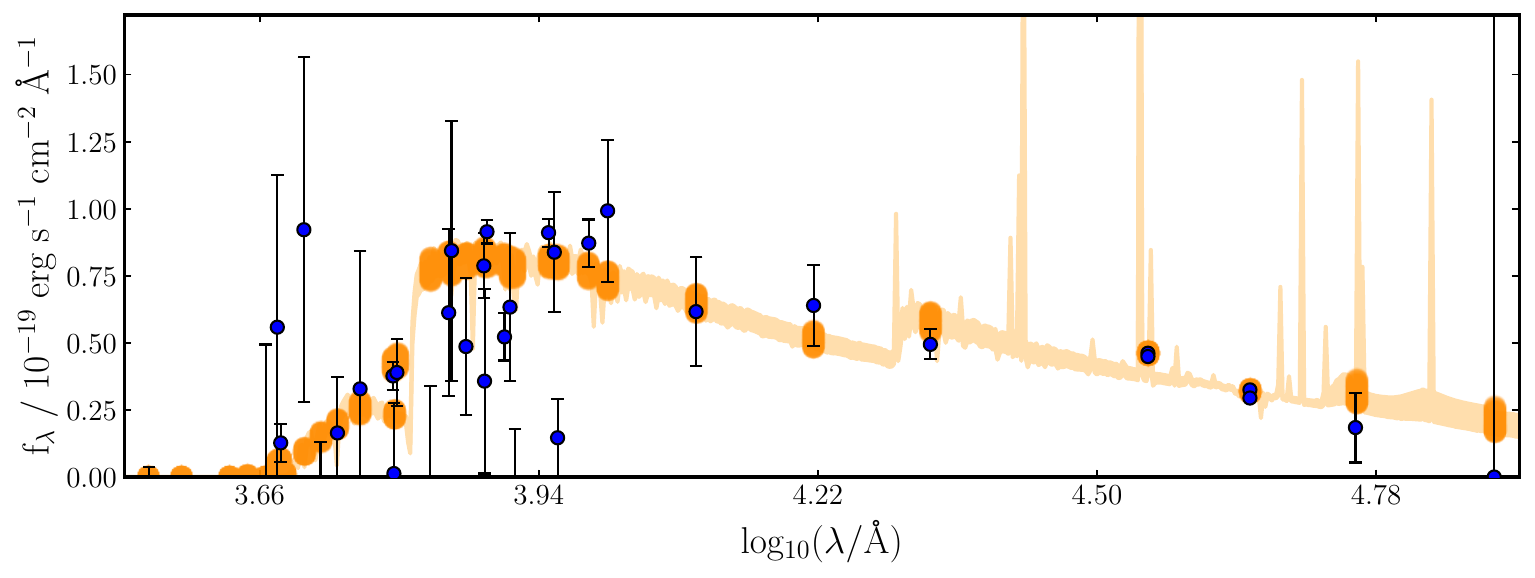}
        \put(47.5,32){\textbf{353556}}
    \end{overpic}
    \begin{overpic}[width=0.48\linewidth]{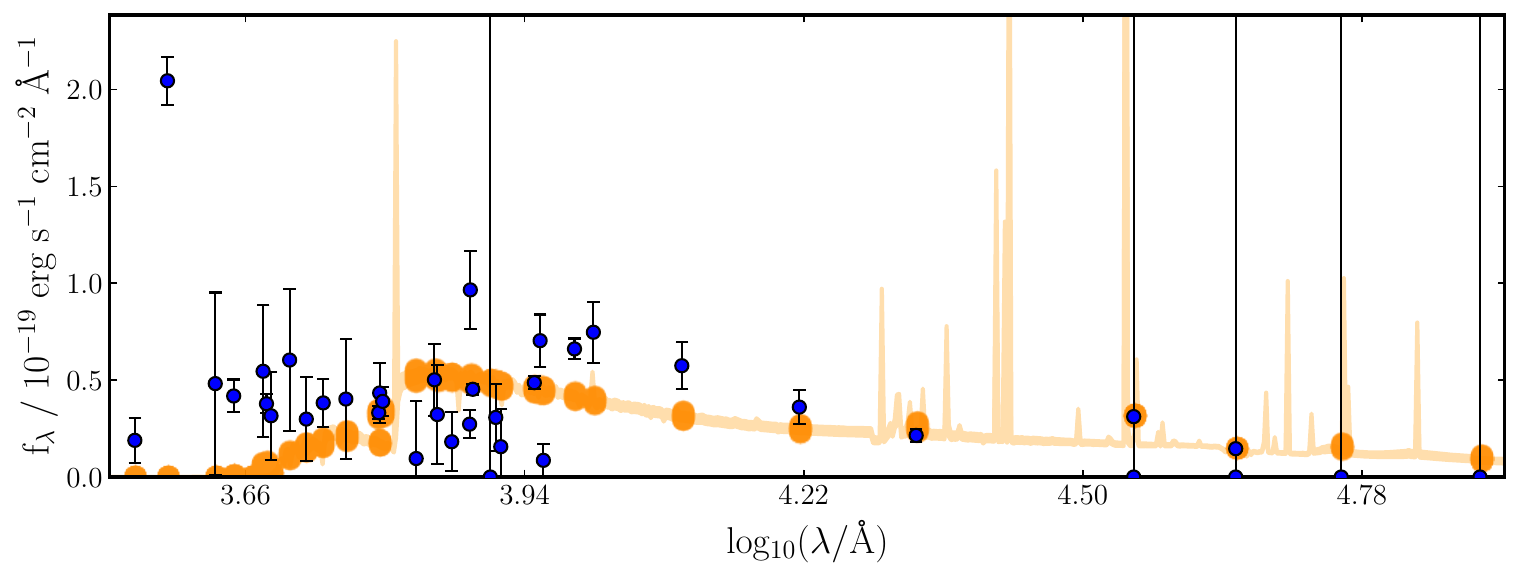}
        \put(47.5,32){\textbf{354768}}
    \end{overpic}
    \begin{overpic}[width=0.48\linewidth]{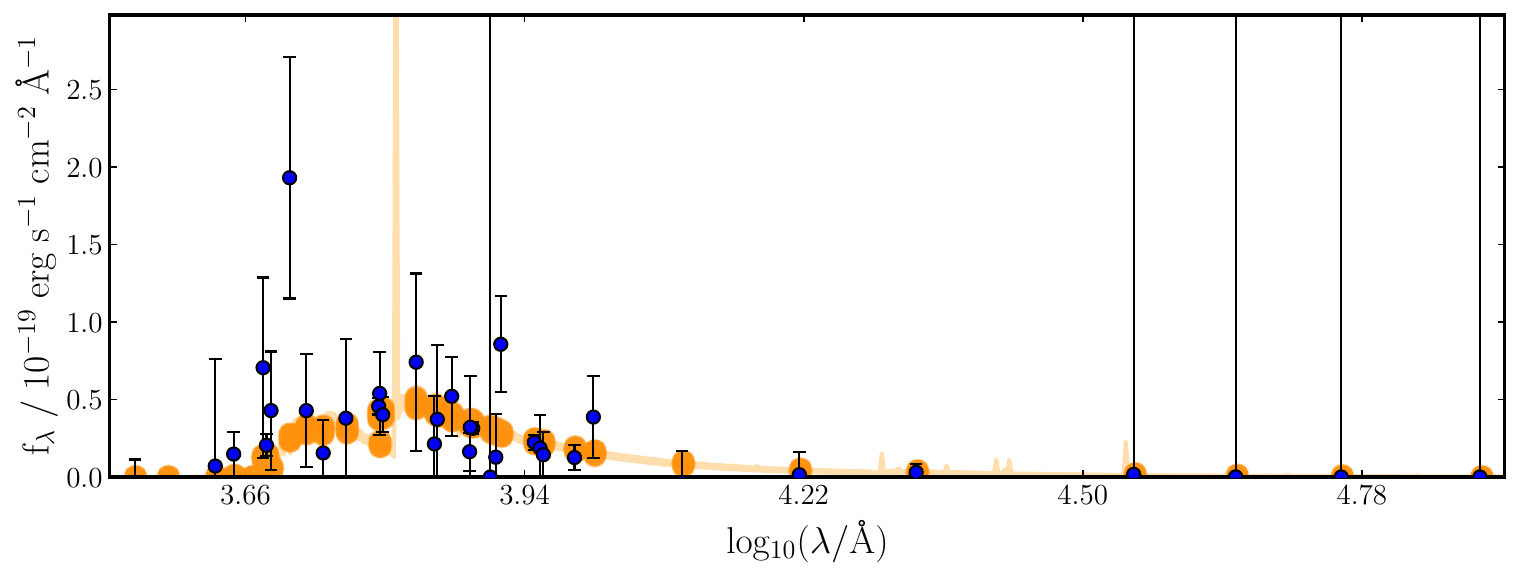}
        \put(47.5,32){\textbf{355293}}
    \end{overpic}    
    \begin{overpic}[width=0.48\linewidth]{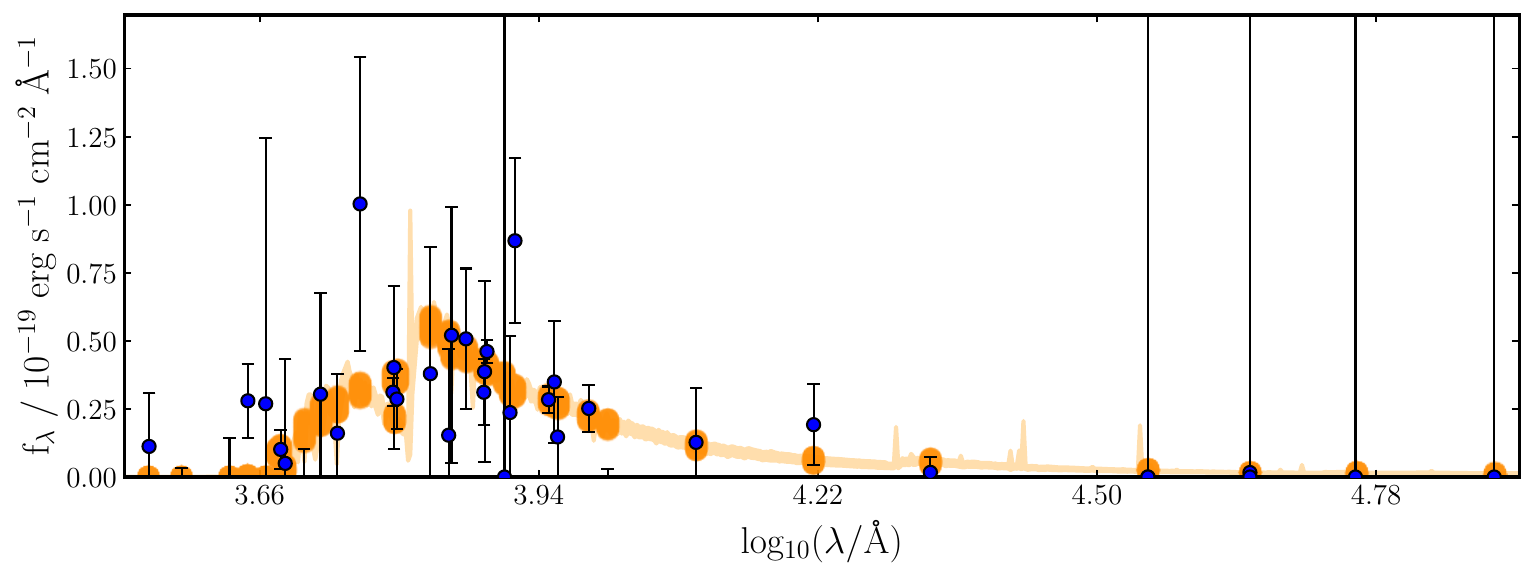}
        \put(47.5,32){\textbf{358735}}
    \end{overpic} 
    \begin{overpic}[width=0.48\linewidth]{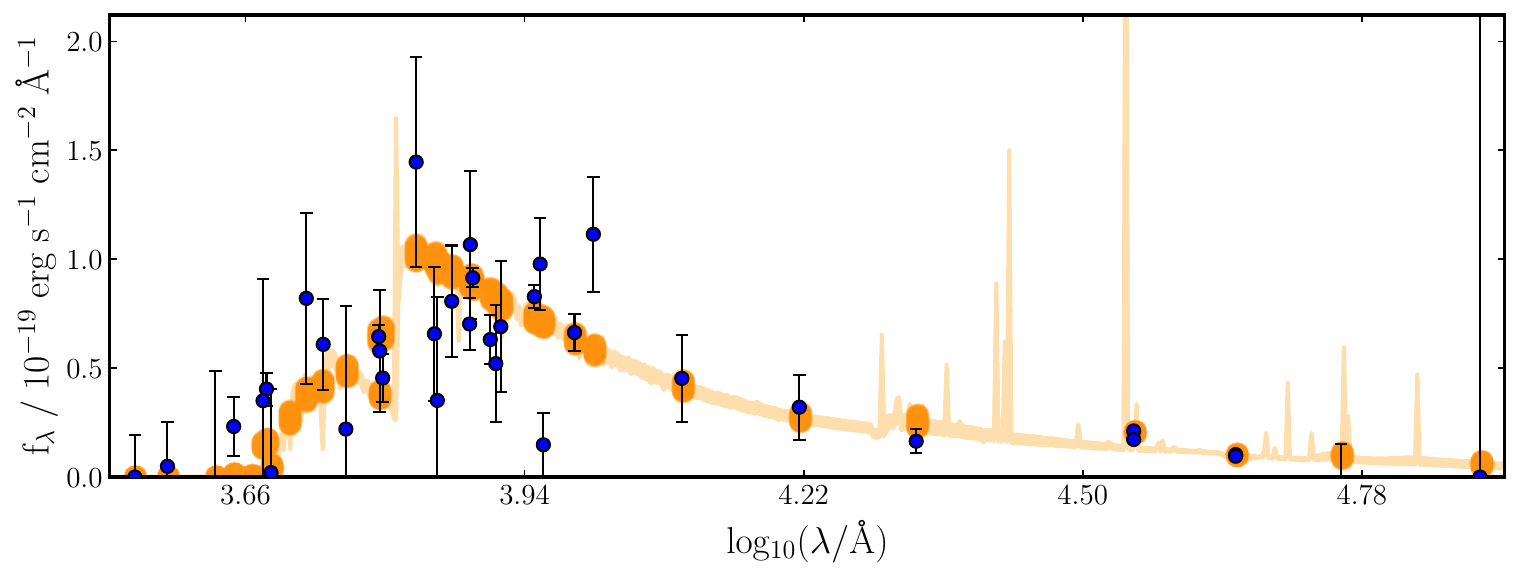}
        \put(47.5,32){\textbf{358920}}
    \end{overpic}     
    \begin{overpic}[width=0.48\linewidth]{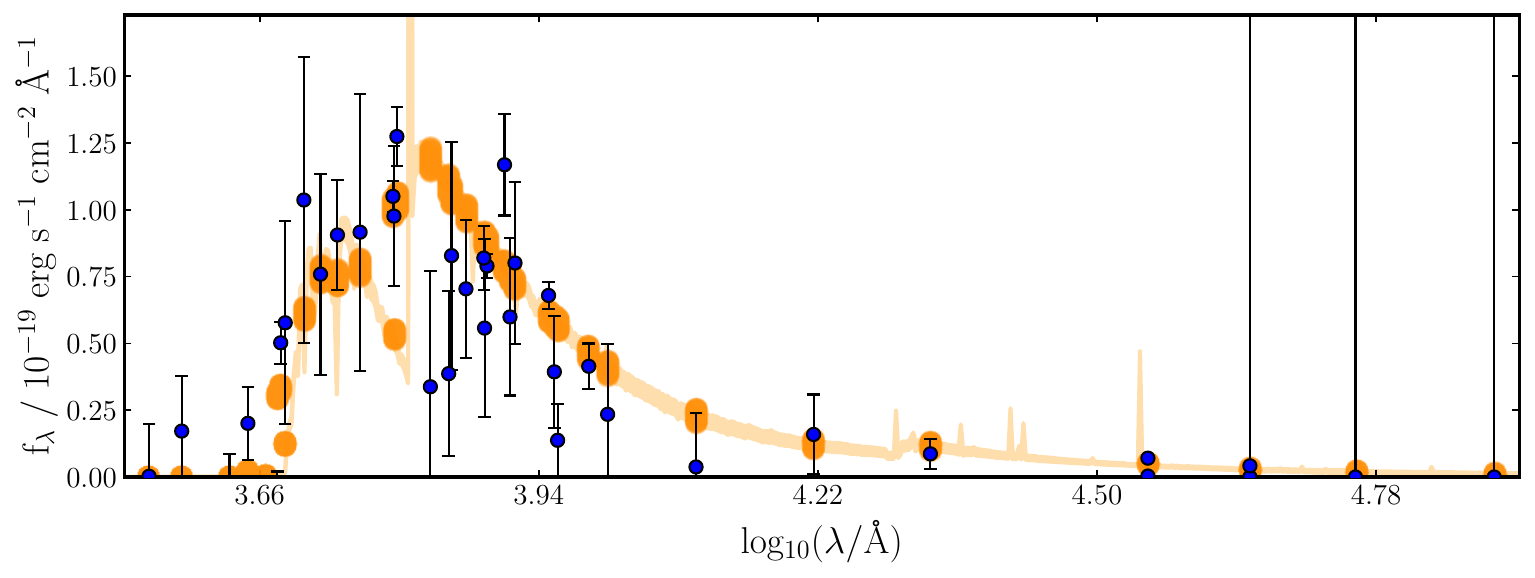}
        \put(47.5,32){\textbf{361346}}
    \end{overpic} 
    \begin{overpic}[width=0.48\linewidth]{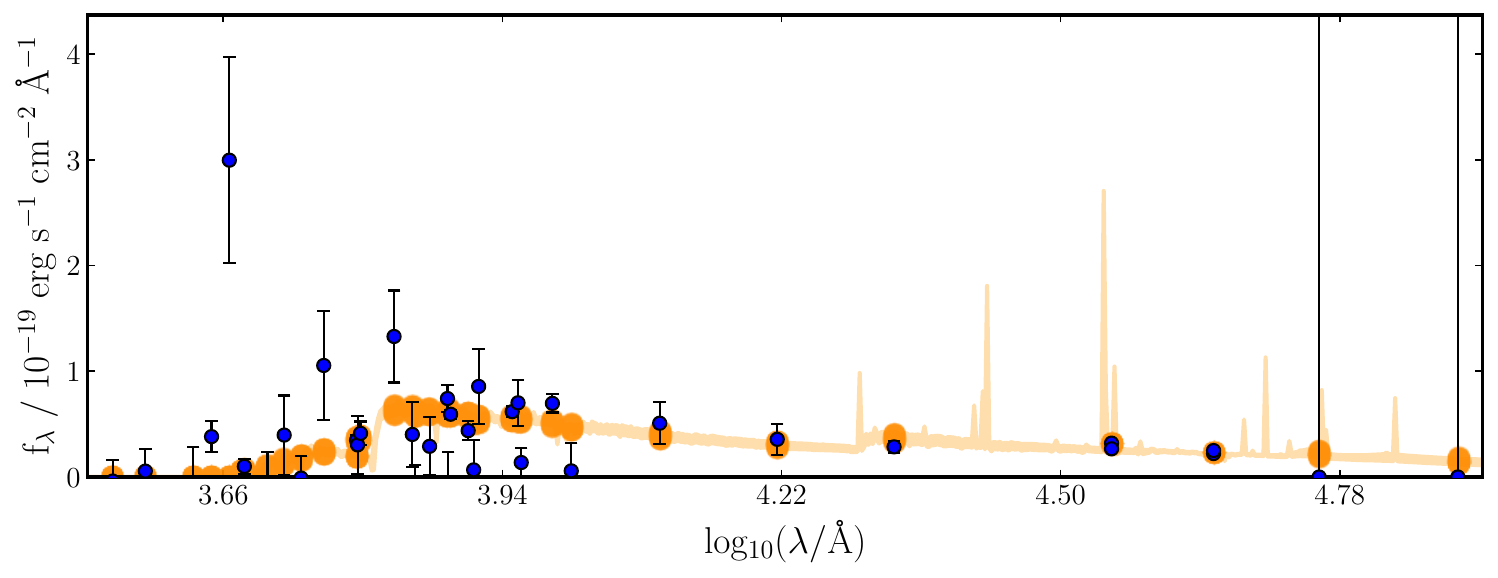}
        \put(47.5,32){\textbf{362449}}
    \end{overpic} 
    \begin{overpic}[width=0.48\linewidth]{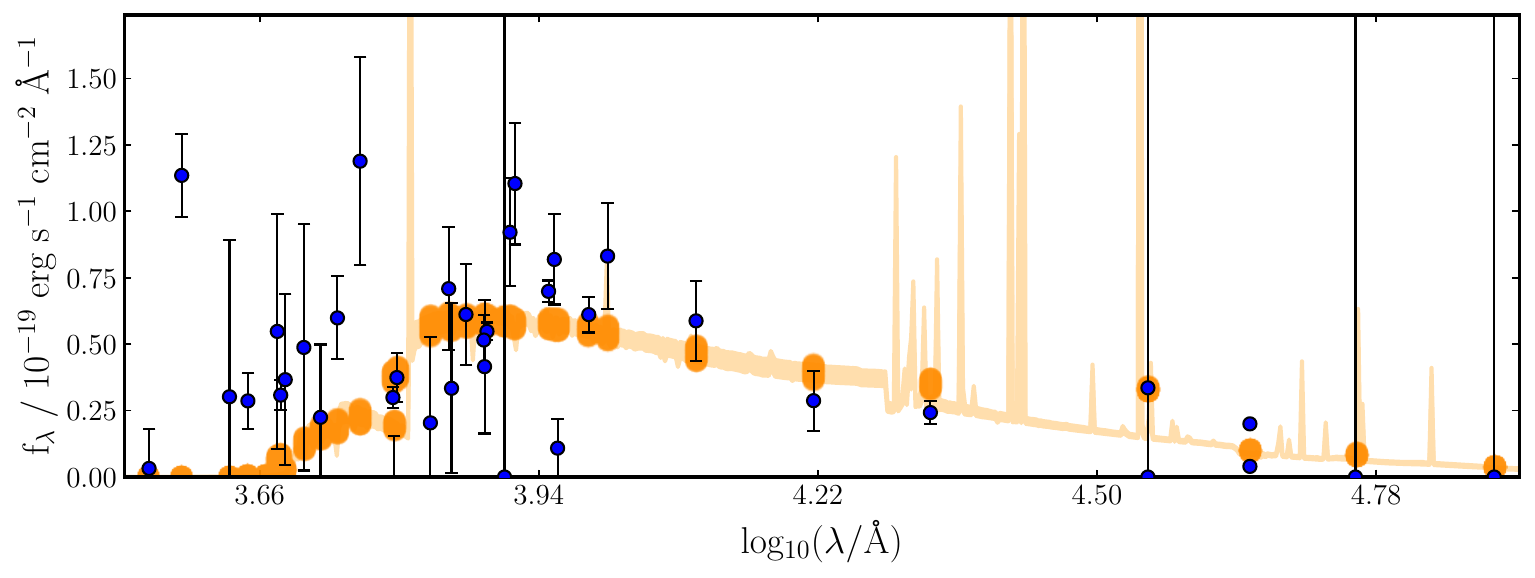}
        \put(47.5,32){\textbf{353918}}
    \end{overpic} 
    \begin{overpic}[width=0.48\linewidth]{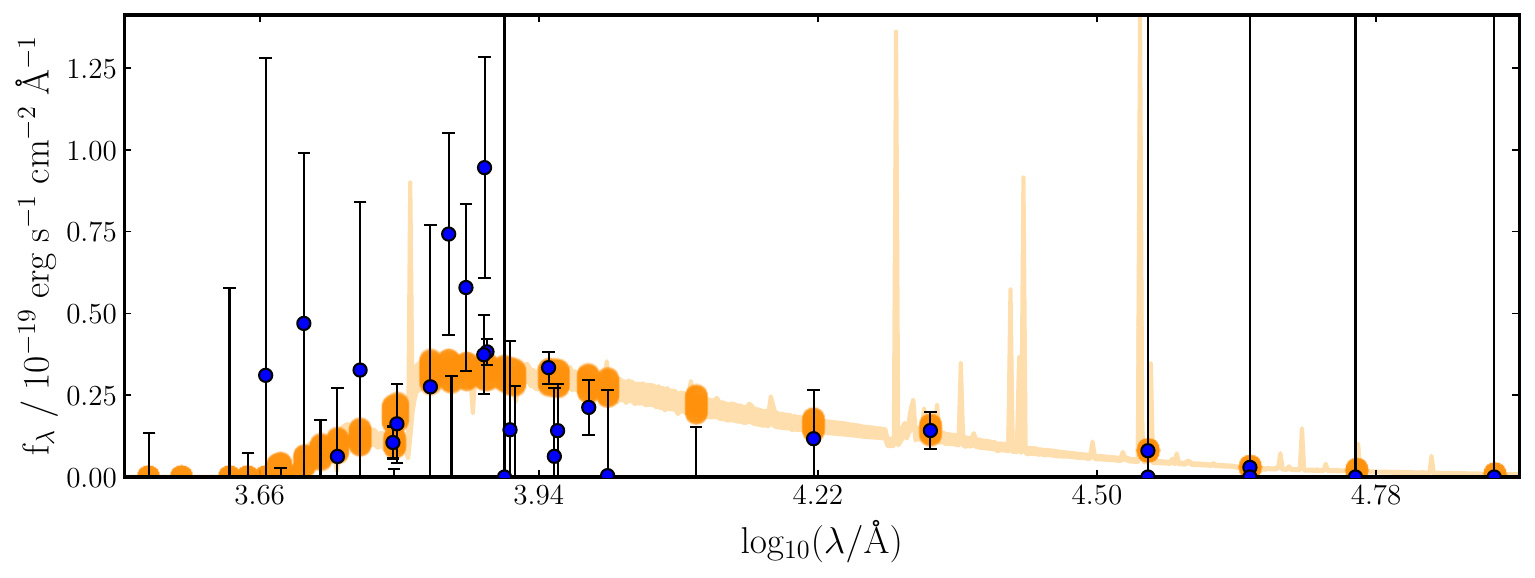}
        \put(47.5,32){\textbf{357448}}
    \end{overpic}
    \begin{overpic}[width=0.48\linewidth]{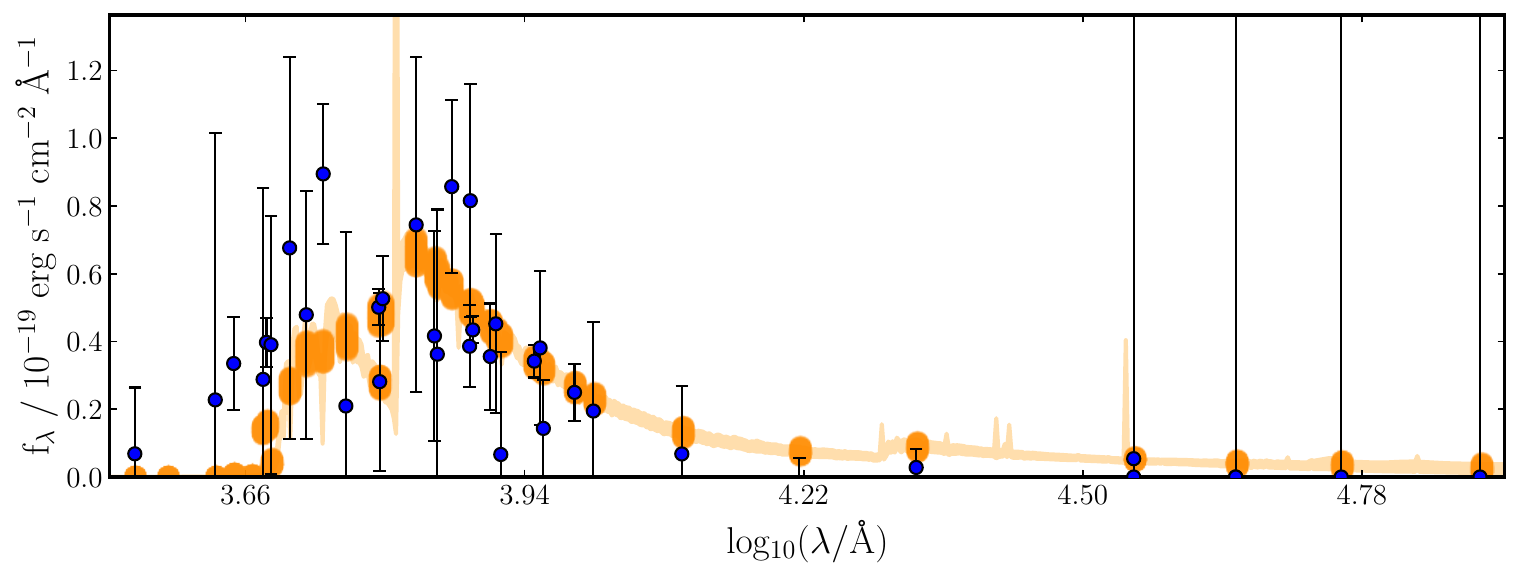}
        \put(47.5,32){\textbf{355779}}
    \end{overpic}
    \caption{Bagpipes SED fits of the optical/NIR detected galaxies in CGG-z4. SEDs were fit at $z=4.331$, the redshift of CGG-z4.a.}
    \label{fig:BPSEDS}
\end{figure}
\begin{figure}[h]\ContinuedFloat
    \centering
    \begin{overpic}[width=0.48\linewidth]{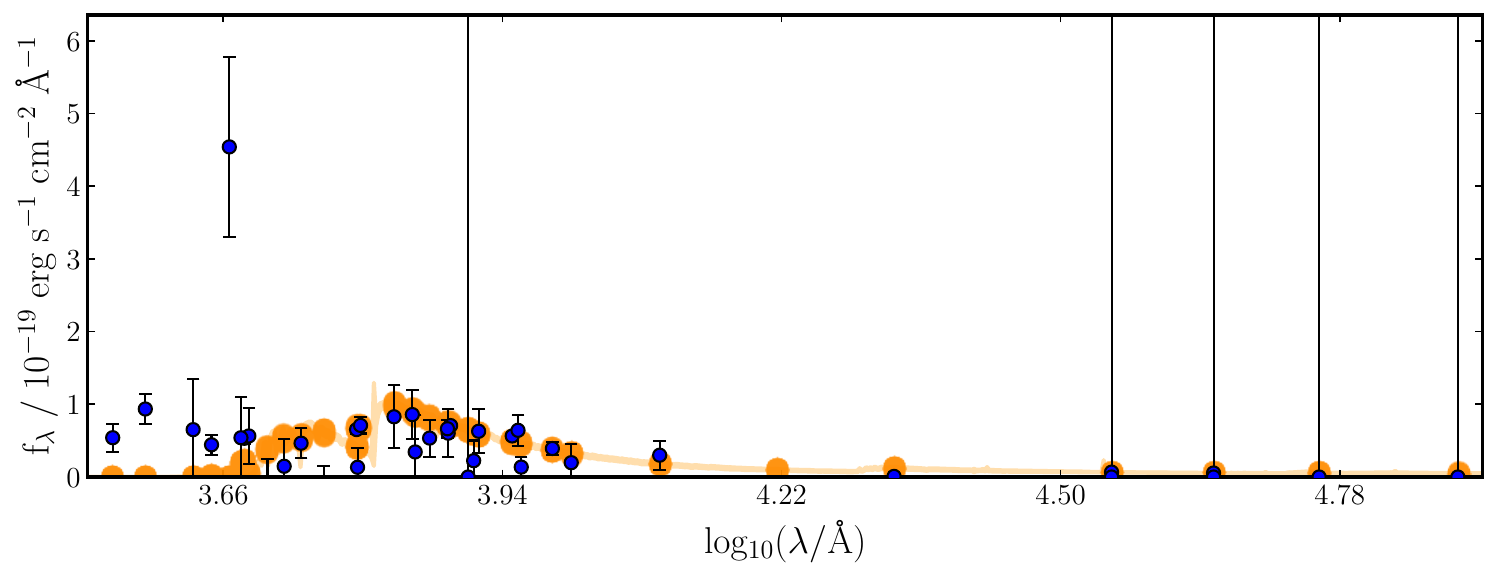}
        \put(47.5,32){\textbf{361388}}
    \end{overpic}
    \caption{Continued.}
    \label{fig:BPSEDS2}
\end{figure}

\section{Galaxy properties from SED fitting}
\begin{table}[h]
\caption[\protect]{SED fitting result for CGG-$z4$ group galaxies.
}
\label{tab:SED}
\centering
\renewcommand*{\arraystretch}{1.5}
\begin{tabular}{l l l l l l l}
\hline
\hline
ID(COSMOS) & $z$ & $\log {\rm M_{\rm \star}/M_{\rm \odot}}$ & SFR & $\rm A_{\rm V}$ & Z & Age\\ 
 &  & [dex] & [M$_{\rm \odot}$ yr$^{-1}$] & & [Z$_{\rm \odot}$] & [Gyr] \\
(1) & (2) & (3) & (4) & (5) & (6) & (7)\\  
\hline
Spec-z\\
\hline
CGG-z4.a & 4.331 & 10.77$^{+0.51}_{-0.49}$ & 1408$\pm86$ &  &  & \\
CGG-z4.b & 4.324 & 10.56$^{+0.42}_{-0.26}$ & 703$\pm65$ &  &  & \\
\hline
Photo-z\\
\hline
353556 & 4.44$^{+0.23}_{-0.15}$ & 10.33$^{+0.15}_{-0.27}$ & 241$^{+117}_{-114}$ & 1.29$^{+0.27}_{-0.31}$ & 0.07$^{+0.32}_{-0.05}$ & 0.02$^{+0.01}_{-0.01}$ \\
354768 & 4.37$^{+0.06}_{-0.03}$ & 9.81$^{+0.15}_{-0.20}$ & 71$^{+31}_{-27}$ & 1.31$^{+0.74}_{-0.44}$ & 0.05$^{+0.08}_{-0.04}$ & 0.01$^{+0.01}_{-0.01}$\\
355293 & 3.87$^{+0.20}_{-0.24}$ & 8.51$^{+0.65}_{-0.71}$ & 2$^{+4}_{-1}$ & 0.57$^{+1.22}_{-0.40}$ & 0.04$^{+0.07}_{-0.02}$ & 0.05$^{+0.22}_{-0.04}$\\
358735 & 4.54$^{+0.11}_{-0.16}$ & 9.28$^{+0.53}_{-0.37}$ & 12$^{+27}_{-7}$ & 1.37$^{+1.35}_{-0.90}$ & 0.53$^{+0.74}_{-0.41}$ & 0.10$^{+0.17}_{-0.07}$\\
358920 & 4.35$^{+0.13}_{-0.20}$ & 9.75$^{+0.12}_{-0.16}$ & 64$^{+20}_{-20}$ & 1.11$^{+0.38}_{-0.29}$ & 0.03$^{+0.06}_{-0.02}$ & 0.01$^{+0.01}_{-0.01}$\\
361346 & 3.90$^{+0.18}_{-0.17}$ & 9.43$^{+0.63}_{-0.17}$ & 6$^{+21}_{-2}$ & 0.46$^{+1.71}_{-0.32}$ & 0.02$^{+0.02}_{-0.01}$ & 0.38$^{+0.08}_{-0.13}$\\
%361481 & 4.80$^{+0.12}_{-0.14}$ & 9.41$^{+0.70}_{-0.53}$ & 16$^{+64}_{-10}$ & 0.94$^{+1.68}_{-0.69}$ & 0.43$^{+0.76}_{-0.36}$ & 0.08$^{+0.13}_{-0.05}$\\
362449 & 4.33$^{+0.18}_{-0.27}$ & $10.18^{+0.27}_{-0.25}$ & $148^{+159}_{-56}$ & $1.52^{+0.52}_{-0.41}$ & $0.20^{+0.79}_{-0.16}$ & $0.02^{+0.04}_{-0.01}$\\
353918 & 4.51$^{+0.16}_{-0.12}$ & $9.20^{+0.08}_{-0.11}$ & $17^{+4}_{-4}$ & $0.36^{+0.18}_{-0.10}$ & $0.02^{+0.01}_{-0.01}$ & $0.01^{+0.01}_{-0.01}$ \\
357448 & 4.62$^{+0.10}_{-0.08}$ & $8.88^{+0.13}_{-0.21}$ & $9^{+3}_{-4}$ & $0.19^{+0.16}_{-0.09}$ & $0.05^{+0.28}_{-0.04}$ & $0.03^{+0.02}_{-0.01}
$ \\
355779 & 4.02$^{+0.15}_{-0.26}$ & $10.19^{+0.42}_{-0.59}$ & $41^{+73}_{-31}$ & $2.53^{+0.99}_{-1.33}$ & $0.03^{+0.05}_{-0.01}$ & $0.31^{+0.14}_{-0.18}$ \\
361388 & 4.23$^{+0.09}_{-0.09}$ & 10.44$^{+0.23}_{-0.23}$ & $115^{+121}_{-55}$ & $3.01^{+0.58}_{-0.71}$ & $0.03^{+0.06}_{-0.02}$ &  $0.19^{+0.20}_{-0.12}$\\
\hline
Total \\
\hline
CGG-z4 & 4.30$^{+0.12}_{-0.14}$ & 11.29$^{+0.38}_{-0.33}$ & 2837$^{+731}_{-472}$ & & & \\
\hline
\end{tabular}
\tablefoot{CGG-z4.a,b were fitted using the {\sc MICHI2} SED fitting code of \cite{Liu2020,Liu2021} to determine the SFR. For galaxies with photometric redshift, results are the median photometric redshift from {\sc EAZY} SED fitting with Classic COSMOS2020 photometry, and the remaining parameters were computed by {\sc Bagpipes} at the spectroscopic redshift of CGG-z4.a, assuming a constant SFH and a \cite{Salim2018} attenuation curve. Columns: (1) Name/COSMOS2020 catalog ID; (2) Redshift; (3) The stellar mass; (4) The star formation rate; (5) Reddening using a \cite{Salim2018} dust law; (6) Metallicity; (7) Mass-weighted age.}
\end{table} 
\clearpage

\section{Gas mass estimates and their products}
\begin{table}[h]
\caption[\protect]{Gas mass estimates for CGG-z4.a,b, and the products calculated from them.}
\label{tab:Gasvalue}
\centering
\renewcommand*{\arraystretch}{1.5}
\begin{tabular}{l l l l l l}
\hline
\hline
ID & M$_{\rm gas,thin}$ & M$_{\rm gas,thick}$ & M$_{\rm gas,\alpha_{CO}=0.8}$ & M$_{\rm gas,\alpha_{CO}=3.8}$ & M$_{\rm gas,CI}$\\ 
(1) & (2) & (3) & (4) & (5) & (6)\\  
\hline
Gas mass  & log([M$_{\rm \odot}$])  & log([M$_{\rm \odot}$]) & log([M$_{\rm \odot}$]) & log([M$_{\rm \odot}$]) & log([M$_{\rm \odot}$]) \\
\hline
CGG-z4.a & $11.29^{+0.12}_{-0.16}$ & $11.07^{+0.15}_{-0.24}$ & $10.51^{+0.04}_{-0.04}$ & $11.05^{+0.04}_{-0.04}$ & $11.14^{+0.09}_{-0.11}$ \\
CGG-z4.b & $10.99^{+0.12}_{-0.17}$ & $10.77^{+0.15}_{-0.24}$  & $10.28^{+0.05}_{-0.06}$ & $10.82^{+0.05}_{-0.06}$ & $<10.66$\\
\hline
Gas mass ratio ($\rm M_{\rm gas}/M_{\rm \star}$) &  & & & \\
\hline
CGG-z4.a & $1.7^{+3.6}_{-0.9}$ & $1.0^{+2.2}_{-0.6}$ & $0.3^{+0.6}_{-0.1}$ & $1.0^{+2.1}_{-0.5}$ & $1.2^{+2.5}_{-0.6}$\\
CGG-z4.b & $1.7^{+3.4}_{-0.9}$ & $1.0^{+2.1}_{-0.6}$ & $0.3^{+0.6}_{-0.2}$ & $1.2^{+2.3}_{-0.6}$ & $<2.3$\\
\hline
Gas depletion time ($\rm M_{\rm gas}/SFR$) & [Myr] & [Myr] & [Myr] & [Myr] & [Myr]\\
\hline
CGG-z4.a & $140^{+44}_{-44}$ & $84^{+36}_{-36}$ & $23^{+2}_{-2}$ & $80^{+9}_{-9}$ & $99^{+23}_{-23}$ \\
CGG-z4.b & $140^{+46}_{-46}$ & $84^{+37}_{-37}$ & $27^{+4}_{-4}$ & $94^{+15}_{-15}$ & $<69$\\
\hline
\end{tabular}
\tablefoot{Columns: (1) shows the name of the source and (2) gas mass using thin dust model; (3) gas mass using thick dust model; (4) gas mass using an $\alpha_{\rm CO}$ numerical value of 0.8; (5) gas mass using an $\alpha_{\rm CO}$ numerical value of $3.8\pm0.1$; (6) gas mass using an $\alpha_{\rm CI}$ numerical value of $16.2\pm0.4$. Due to the large uncertainty on the stellar masses, the gas mass ratios are estimated through Monte Carlo sampling, generating $10^{6}$ samples and taking their median and $68\%$ confidence interval.}
\end{table}

\end{appendix}

\end{document}